\documentclass[journal]{IEEEtran}
\ifCLASSINFOpdf
\else
\fi

\usepackage{cite}
\usepackage{physics,amsmath,amssymb,amsfonts}
\usepackage[ruled, lined, linesnumbered, commentsnumbered, longend]{algorithm2e}
\SetCommentSty{mycommfont}
\usepackage{floatrow}
\usepackage{graphicx}
\usepackage{textcomp}
\usepackage{xcolor}
\usepackage{subcaption}
\usepackage{enumitem}
\usepackage{makecell}
\usepackage[normalem]{ulem}

\usepackage{lipsum,adjustbox}
\usepackage{tikz}
\usetikzlibrary{decorations.pathreplacing}

\def\BibTeX{{\rm B\kern-.05em{\sc i\kern-.025em b}\kern-.08em
    T\kern-.1667em\lower.7ex\hbox{E}\kern-.125emX}}
\usepackage{enumitem}
\usepackage{xcolor}

\usepackage{url,hyperref}
\usepackage[linguistics]{forest}

\begin{document}

\title{Generative AI and Federated Learning for Intrusion Detection Systems: A Survey}

\author{Jiefei Liu, Abu Saleh Md Tayeen, Pratyay Kumar, Qixu Gong, Wenbin Jiang, \\ Huiping Cao, Satyajayant Misra, Jayashree Harikumar
\thanks{
Jiefei Liu, Pratyay Kumar, Qixu Gong, Wenbin Jiang, Huiping Cao, and Satyajayant Misra are with the Department of Computer Science, New Mexico State University, Las Cruces, NM, USA (e-mail: \{jiefei, pratyay, qixugong, wbjiang, hcao, misra\}@nmsu.edu).
}
\thanks{
Abu Saleh Md Tayeen is with the University of Hartford, CT, USA (e-mail:tayeen@hartford.edu).
}
\thanks{Jayashree Harikumar is with DEVCOM Analysis Center, WSMR, NM, USA (e-mail: jayashree.harikumar.civ@army.mil).}
}

\maketitle

\begin{abstract}
Intrusion Detection Systems (IDSs) are essential for monitoring network traffic and identifying malicious activities in modern cyber-physical, Internet of Things (IoT), enterprise, and distributed network environments. However, developing reliable IDS models remains challenging because attack behaviors evolve over time, realistic datasets are difficult to obtain, traffic records may be incomplete, attack classes are often imbalanced, and privacy constraints limit centralized data collection. Recent advances in generative artificial intelligence (AI) and Federated Learning (FL) provide new opportunities to address these limitations. Generative models can support anomaly detection, synthetic traffic generation, data augmentation, data imputation, adversarial traffic generation, and IDS alert explanation. FL enables distributed IDS training without directly sharing local network traffic, making it suitable for privacy-sensitive and geographically distributed environments.

This survey provides a structured review of generative AI and FL techniques for IDS. We first summarize representative IDS research directions, including adversarial machine learning, anomaly-based detection, IoT-oriented IDS, explainable IDS, and benchmark datasets. We then categorize generative AI applications in IDS according to model families and task objectives, covering autoencoder-based models, Generative Adversarial Networks (GANs), diffusion models, and Large Language Models (LLMs). Finally, we review emerging studies that integrate generative AI with FL-based IDS and discuss open challenges, including synthetic data quality, realistic traffic generation, dual-use adversarial risks, non-IID client distributions, communication-efficient model sharing, federated IDS benchmarking, and domain-specific LLMs for network security.
\end{abstract}


\IEEEpeerreviewmaketitle

\section{Introduction} \label{sec:introduction}

Modern computer networks support a wide range of critical services, including cloud computing, Internet of Things (IoT) platforms, industrial control systems, intelligent transportation, financial systems, healthcare infrastructure, and defense applications. As these systems become increasingly connected, they also become more exposed to cyber attacks that can disrupt services, compromise sensitive data, or damage physical infrastructure. Intrusion Detection Systems (IDSs) are therefore an essential component of network security. An IDS monitors system activities or network traffic and identifies behaviors that may indicate unauthorized access, malware propagation, denial-of-service attacks, data exfiltration, or other malicious activities.
Early IDS research established the foundation for monitoring security-relevant events and detecting abnormal system behaviors~\cite{anderson1980computer,DBLP:journals/tse/Denning87}.
Since then, IDS techniques have evolved from rule-based and signature-based detection toward data-driven approaches based on machine learning (ML) and deep learning (DL). ML-based IDS models can learn complex traffic patterns from historical data and have shown strong performance in both binary attack detection and multi-class attack classification. However, their effectiveness strongly depends on the availability, quality, diversity, and representativeness of training data. In practice, network intrusion datasets are often limited, imbalanced, incomplete, outdated, or collected from restricted environments that do not fully reflect real-world network conditions.

Generative artificial intelligence (AI) provides a promising direction for addressing these data-related limitations. Generative models learn the underlying distribution or structure of observed data and can produce new samples that resemble the original data. In the IDS domain, generative models have been used for synthetic traffic generation, data augmentation, missing-value imputation, anomaly detection, adversarial traffic generation, and explanation of IDS alerts. Representative generative techniques include autoencoders and variational autoencoders, generative adversarial networks, diffusion models, and large language models. These methods offer new opportunities to improve IDS robustness, especially when real attack samples are scarce, minority classes are underrepresented, or data collection is expensive.

At the same time, the deployment of IDS models faces increasing privacy and communication constraints. Traditional centralized IDS training requires collecting network traffic from distributed clients or organizations and transferring it to a central server. This strategy can expose sensitive information and is often impractical for privacy-sensitive environments. 
Federated Learning (FL) addresses this issue by allowing clients to train models locally and share only model updates with a central server~\cite{DBLP:conf/aistats/McMahanMRHA17}.
For IDS, FL is particularly attractive because network traffic is naturally distributed across devices, routers, organizations, and geographic regions. However, FL-based IDS also introduces new challenges, including non-independent and identically distributed (non-IID) client data, communication overhead, client heterogeneity, poisoning attacks, and limited access to realistic FL-based IDS benchmarks.

Although IDS, generative AI, and FL have each been studied extensively, their intersection remains fragmented. Existing studies often focus on a specific model family, a single IDS task, or an isolated FL setting. A systematic review is needed to clarify how generative models are used in IDS, how they can support FL-based IDS, and what technical challenges remain unresolved. Motivated by this need, this survey reviews recent progress in generative AI for IDS and generative AI-embedded FL-based IDS. We organize the literature by IDS research problems, generative model families, application objectives, and FL integration strategies, and we further discuss open challenges and future research directions for building reliable, privacy-preserving, and data-efficient IDS models.

\subsection{Research Problems}\label{subsec:research_problem}

Although IDS has been widely studied, the integration of generative AI and Federated Learning (FL) introduces several open research problems that remain scattered across the literature. This survey focuses on the following questions.

\textbf{RQ1: How are generative models used to improve IDS?}
Generative models have been applied to IDS for multiple purposes, including anomaly detection, synthetic traffic generation, data augmentation, data imputation, adversarial traffic generation, and alert explanation. However, these applications are often studied independently. A systematic review is needed to clarify which generative model families are used, what IDS problems they address, and how their assumptions differ.

\textbf{RQ2: How reliable is synthetic network traffic for IDS training and evaluation?}
Synthetic data can mitigate limited data availability, class imbalance, and missing records. However, high statistical similarity to real data does not necessarily guarantee realistic network behavior. In IDS, generated traffic should also preserve protocol constraints, temporal dependencies, attack semantics, and network-topology relationships. Therefore, evaluating the quality, usefulness, and realism of synthetic IDS data remains a key challenge.

\textbf{RQ3: How can generative AI support privacy-preserving and communication-efficient FL-based IDS?}
FL allows distributed clients to train IDS models without directly sharing raw traffic data. However, FL-based IDS faces non-IID client distributions, communication overhead, client heterogeneity, and vulnerability to poisoning attacks. Generative AI can potentially address these issues by augmenting local data, improving minority-class representation, reducing data heterogeneity, or generating privacy-preserving synthetic samples. The effective integration of generative models into FL-based IDS is still an emerging research direction.

\textbf{RQ4: What datasets and benchmarks are available for evaluating IDS, generative IDS, and FL-based IDS?}
Reliable evaluation depends on realistic and representative datasets. However, real network traffic is often sensitive and difficult to release publicly. Existing IDS datasets are useful but may be outdated, centrally collected, or insufficient for evaluating federated and generative settings. This survey therefore summarizes commonly used IDS datasets and discusses the need for more realistic benchmarks for generative AI and FL-based IDS.

Based on these research problems, this survey reviews IDS background studies, categorizes generative AI techniques for IDS, summarizes existing IDS datasets, and discusses emerging directions in generative AI-embedded FL-based IDS.

%=============================
\begin{table}[htbp]
\renewcommand{\arraystretch}{1.5} % Default value: 1
\begin{tabular}{|c | l }
\multicolumn{2}{|l}{\textbf{Section~\ref{sec:introduction} Introduction}} \\
  & Section~\ref{subsec:research_problem} Research problems \\ 

  & Section~\ref{subsec:contributions} Contributions \\  

  & Table~\ref{tab:paper_structure} Structure of our survey \\  

\multicolumn{2}{|l}{\textbf{Section~\ref{sec:IDS} Intrusion detection systems}} \\

  & Section~\ref{subsec:aml-ids} Adversarial Machine Learning for IDS \\ 

  & Section~\ref{subsec:anomal-ids} Anomaly-based IDS \\  

  & Section~\ref{subsec:ids-iot} IDS for IoT\\  

  & Section~\ref{subsec:explainable_IDS} Explainable IDS (X-IDS)\\  

  & Section~\ref{subsec:dataset-ids} Datasets for IDS\\  

\multicolumn{2}{|l}{\textbf{Section~\ref{sec:generative_models} Generative models}} \\

  & Section~\ref{subsec:VAE_IDS} Variational Autoencoder \\ 
  
  & Section~\ref{subsec:GAN_IDS} GAN in IDS \\  

  & Section~\ref{subsec:Diffusion_IDS} Diffusion in IDS \\  

  & Section~\ref{subsec:LLM_in_IDS} LLM in IDS\\  

\multicolumn{2}{|l}{\textbf{\makecell{Section~\ref{sec:AI_FL_IDS} Generative AI Embedded Federated\\ Learning based Intrusion Detection System}}} \\

  & Section~\ref{subsec:VAE_FL_IDS} VAE embedded FL-IDS\\ 

  & Section~\ref{subsec:GAN_FL_IDS} GAN embedded FL-IDS\\  

  & Section~\ref{subsec:Diffusion_FL_IDS} Diffusion embedded FL-IDS\\  

  & Section~\ref{subsec:LLM_FL_IDS} LLM embedded FL-IDS \\

\multicolumn{2}{|l}{\textbf{Section~\ref{sec:conclusion} Conclusion}} \\
\end{tabular}
\caption{Structure of this survey}
\label{tab:paper_structure}
\end{table}
%=============================

\subsection{Contributions}\label{subsec:contributions}

The main contributions of this survey are summarized as follows:

\begin{itemize}
    \item We provide a structured review of IDS research from the perspective of data-driven security modeling, covering representative directions including adversarial machine learning for IDS, anomaly-based IDS, IoT-oriented IDS, explainable IDS, and IDS benchmark datasets. This background establishes the technical context for understanding why generative AI and Federated Learning (FL) are increasingly important for modern IDS development.

    \item We present a taxonomy of generative AI applications in IDS by organizing existing studies according to both model family and task objective. Specifically, we review autoencoder-based models, Generative Adversarial Networks (GANs), diffusion models, and Large Language Models (LLMs), and analyze how they are used for anomaly detection, synthetic traffic generation, data augmentation, data imputation, adversarial traffic generation, and IDS alert explanation.

    \item We systematically examine the emerging intersection of generative AI and FL-based IDS. Different from conventional IDS surveys that study centralized detection models, this survey highlights how generative models can support privacy-preserving and distributed IDS training by addressing non-IID client data, class imbalance, communication cost, adversarial robustness, and limited local data availability.

    \item We summarize commonly used IDS datasets and FL-oriented IDS datasets, emphasizing their roles, limitations, and suitability for evaluating generative AI and FL-based IDS. This dataset-level discussion helps clarify current benchmark gaps, especially the shortage of realistic, topology-aware, and federated IDS datasets.

    \item We identify open challenges and future research directions for generative AI-enabled IDS, including synthetic data reliability, realistic network traffic generation, evaluation of generated samples, privacy-preserving data augmentation, communication-efficient generative FL, and domain-specific LLMs for network security.
\end{itemize}

\subsection{Survey Methodology}
To assemble the literature reviewed in this survey, we conducted a keyword-based search across Google Scholar and DBLP, supplemented by manual inspection of recent proceedings and articles from major security, networking, and machine learning venues. Our search combined two sets of terms: generative-model keywords (generative AI, VAE, variational autoencoder, GAN, diffusion model, large language model, LLM) and application keywords (intrusion detection, IDS, network security, federated learning, anomaly detection, synthetic traffic, data augmentation). Candidate papers were then selected primarily by relevance rather than by a fixed time window: a study was included if it applied a generative model family to an IDS task, integrated generative models with FL-based IDS, or provided foundational background (e.g., seminal generative architectures, IDS datasets, or evaluation metrics) needed to interpret these works. We excluded papers that used generative models in unrelated domains without transferable methodology and those that did not provide enough methodological detail to assess. Because generative AI is a fast-moving field, the large majority of the application-oriented studies we review were published within roughly the last five to seven years, while a smaller set of older references is retained to establish the technical lineage of each model family (for example, the original VAE, GAN, and diffusion formulations) and the early foundations of intrusion detection. This relevance-driven strategy lets the survey track current practice while preserving the conceptual context required to compare model families and identify open challenges.

Following this methodology, the remainder of this survey is organized as shown in Table~\ref{tab:paper_structure}. Section~\ref{sec:IDS} first provides the background of IDS by reviewing representative research directions, including adversarial machine learning, anomaly-based detection, IoT-oriented IDS, explainable IDS, and commonly used IDS datasets. Section~\ref{sec:generative_models} then reviews generative AI techniques for IDS and categorizes existing studies by model family, including Variational Autoencoders (VAEs), Generative Adversarial Networks (GANs), diffusion models, and Large Language Models (LLMs). Section~\ref{sec:AI_FL_IDS} further discusses the integration of generative AI with Federated Learning (FL)-based IDS, focusing on how different generative models can support distributed, privacy-preserving, and data-efficient IDS training. Finally, Section~\ref{sec:conclusion} concludes the survey and summarizes future research opportunities.

%==============================
\section{Intrusion detection systems (IDS)}\label{sec:IDS}
%================================
%----------------------------
Cybersecurity aims to protect computer systems, networks, and data from unauthorized access, misuse, disruption, and damage. As modern services increasingly depend on interconnected infrastructures, detecting malicious activities before they cause serious impact has become a core requirement for network defense. Intrusion Detection Systems (IDSs) address this requirement by monitoring system events or network traffic and identifying behaviors that may indicate security violations.
The concept of intrusion detection was first introduced by James Anderson in the early 1980s~\cite{anderson1980computer}. Building on this foundation, Denning proposed one of the earliest functional IDS models, which formalized intrusion detection as the process of monitoring audit records and identifying abnormal or suspicious system behavior~\cite{DBLP:journals/tse/Denning87}. In general, an IDS can be implemented as a software- or hardware-based security mechanism. Its primary goal is to detect potential attacks, policy violations, or abnormal activities and provide alerts that support timely investigation and response.

%---------------------------------
Over the past few decades, IDS research has evolved from rule-based and statistical methods to machine learning and deep learning approaches. Existing IDS studies differ not only in the detection algorithms they use, but also in the attack assumptions, deployment environments, data sources, and explanation mechanisms they consider. Therefore, before reviewing generative AI and FL-based IDS, it is necessary to summarize the main IDS research directions that provide the technical context for this survey.
In this section, we review representative IDS-related studies from five perspectives. Section~\ref{subsec:aml-ids} discusses adversarial machine learning for IDS, focusing on the vulnerability of ML-based detectors to adversarial manipulation. Section~\ref{subsec:anomal-ids} reviews anomaly-based IDS, where attacks are detected by modeling deviations from normal behavior. Section~\ref{subsec:ids-iot} summarizes IDS studies for IoT environments, where resource constraints and heterogeneous devices introduce additional challenges. Section~\ref{subsec:explainable_IDS} reviews explainable IDS, which aims to make detection results more interpretable for security analysts. Finally, Section~\ref{subsec:dataset-ids} discusses commonly used IDS datasets, which are fundamental for training, evaluating, and comparing IDS models.

%----------------------------
\subsection{Adversarial Machine Learning for IDS}
\label{subsec:aml-ids}
%----------------------------------
Machine learning (ML) and deep learning (DL) models have been widely adopted in IDS because they can learn discriminative patterns from network traffic and detect attacks beyond manually defined signatures. However, the use of ML also introduces a new attack surface. An adversary may intentionally perturb input traffic features or manipulate training data to mislead the detector, causing malicious traffic to be classified as benign or forcing benign traffic to be reported as malicious. These threats are generally studied under adversarial machine learning, which focuses on understanding the vulnerability of ML models and improving their robustness against adversarial manipulation.

For IDS, adversarial robustness is especially important because network attackers can actively adapt their behaviors after observing or probing the detection system. Even small modifications to traffic characteristics may change the prediction of an ML-based IDS while preserving the attack objective. Therefore, researchers have investigated both adversarial attacks against IDS models and defense strategies that make IDS models more stable under adversarial conditions.

Jmila and Khedher~\cite{DBLP:journals/cn/JmilaK22} evaluated the vulnerability of IDS models built with shallow ML classifiers, including Decision Tree, Random Forest, and Logistic Regression. Their experiments on NSL-KDD~\cite{tavallaee2009detailed} and UNSW-NB15~\cite{moustafa2015unsw} show that different adversarial attacks affect classifiers differently, indicating that IDS robustness depends on both the attack strategy and the underlying model architecture. Alotaibi and Rassam~\cite{fi15020062} further surveyed adversarial attacks against ML-based IDS and summarized representative defense strategies. These studies show that adversarial machine learning is an important research direction for IDS, particularly as IDS models become more data-driven and are deployed in adaptive threat environments.

%--------------------------
\subsection{Anomaly-based IDS} 
\label{subsec:anomal-ids}
%--------------------------

IDSs are commonly categorized as signature-based, anomaly-based, or hybrid systems according to their detection strategy. A signature-based IDS detects intrusions by matching observed activities with predefined attack signatures or rules. This approach is effective for known attacks, but it is limited when facing new or evolving threats that do not match existing signatures. In contrast, an anomaly-based IDS first models normal system or network behavior and then identifies activities that deviate from this learned baseline. This capability makes anomaly-based IDS particularly useful for detecting unknown or zero-day attacks, although it may also introduce higher false-positive rates when benign behavior changes over time.

Several surveys have reviewed anomaly-based IDS from different perspectives. Yang et al.~\cite{DBLP:journals/compsec/YangLLWWZH22} conducted a systematic review of network IDS studies and analyzed commonly used data processing techniques, evaluation metrics, benchmark datasets, and detection models. Hajj et al.~\cite{hajj2021anomaly} presented a taxonomy of network attacks and discussed attack tools, relevant detection features, IDS data sources, dataset types, system architectures, and detection modes. They also summarized key challenges that affect the effectiveness of anomaly-based IDS, including dataset quality, feature selection, evaluation consistency, and deployment constraints.

These studies show that anomaly-based IDS is a central direction in IDS research because it directly addresses the limitation of signature-based detection under unknown attacks. However, its performance depends heavily on how normal behavior is modeled, how representative the training data are, and how deviations are distinguished from benign traffic variations.

%-----------------------
\subsection{IDS for IoT}
\label{subsec:ids-iot}
%------------------------

The Internet of Things (IoT) connects sensors, actuators, embedded devices, and coordinator nodes to provide networked services in domains such as smart homes, healthcare, transportation, and industrial systems. Compared with conventional networks, IoT environments are more heterogeneous because devices may differ in hardware capability, operating system, communication protocol, and deployment context. Many IoT devices also have limited computation, memory, and energy resources, which makes it difficult to deploy complex security mechanisms directly on the devices. In addition, vulnerabilities in firmware, hardware, and protocol implementations can expose IoT systems to malware propagation, denial-of-service attacks, spoofing, and unauthorized access. These characteristics make IDS design for IoT different from traditional IDS design.

Several studies have reviewed IDS techniques specifically for IoT environments. Kumar et al.~\cite{kumar2023comprehensive} provided a taxonomy of ML-based IDS for secure IoT communication and compared different categories according to their advantages, limitations, and resource-related evaluation metrics. They also proposed an IDS model that combines convolutional neural networks (CNNs) with fuzzy rules to improve performance under IoT constraints, including energy consumption and packet delivery ratio. Jayalaxmi et al.~\cite{DBLP:journals/access/JayalaxmiSKCK22} analyzed ML- and DL-based IDS and Intrusion Prevention Systems (IPS) for IoT. They further proposed a risk factor analyzer and a hybrid Intrusion Detection and Prevention System (IDPS) framework to address limitations of purely anomaly-based or signature-based methods.

These studies show that IoT-oriented IDS must consider both detection accuracy and deployment constraints.
Effective IDS for IoT should be able to detect diverse attacks while remaining lightweight, adaptive to heterogeneous devices, and practical for resource-constrained environments.
%-------------------------------
\subsection{Explainable IDS (X-IDS)}\label{subsec:explainable_IDS}
%=========================

Machine learning (ML) and deep learning (DL) models have been widely used in IDS because they can learn complex traffic patterns and improve attack detection performance. However, many DL-based IDS models operate as black-box systems, meaning that their internal decision process is difficult for human analysts to interpret. This lack of transparency limits the ability of administrators and security experts to understand why an alert is generated, identify the root cause of an attack, and determine an appropriate response. Explainable Artificial Intelligence (XAI) addresses this issue by providing interpretable evidence or explanations for model predictions. In the IDS context, explainability is important not only for improving user trust, but also for supporting incident analysis, model debugging, and security decision-making.

Several surveys have studied explainable IDS from different perspectives. Moustafa et al.~\cite{10136827} presented a comprehensive survey on XAI methods for cyber defense, with a particular focus on anomaly-based IDS in IoT networks. Their work reviewed studies at the intersection of XAI, anomaly-based intrusion detection, IoT security, summarized open challenges, and future directions for explainable cyber-defense systems. Neupane \textit{et al.}~\cite{DBLP:journals/access/NeupaneAAMRBS22} proposed a taxonomy of X-IDS techniques and categorized existing methods into white-box and black-box approaches. They also introduced a three-layered X-IDS architecture inspired by the DARPA XAI program~\cite{gunning2019darpa} and discussed key challenges in developing explainable IDS models.

These studies indicate that X-IDS is an important direction for practical IDS deployment. A detection model with high accuracy may still be difficult to use in real security operations if its alerts cannot be interpreted. Therefore, explainable IDS aims to bridge the gap between automated detection and human-centered cyber-defense analysis.

\subsection{Datasets for IDS}
\label{subsec:dataset-ids}
%--------------------------------------

%=============================
\begin{table*}[htbp]
\centering
\begin{tabular}{ c  c  c  c  c  c  c }
\hline
\textbf{Dataset Name} & \textbf{Year} & \textbf{Training instances} & \textbf{Testing instances} & \textbf{\# of classes} &\textbf{\# of Features} & \textbf{FL support}\\ 
\hline
KDD CUP 99~\cite{divekar2018benchmarking} & 1999 & 1,074,992 & 311,029 &  5 & 41 & No\\ 
\hline
 ISCX NSL-KDD~\cite{tavallaee2009detailed} & 2009 & 4,898,431 & 311,027 & 4 & 42 & No\\
\hline
 \textbf{UNSW-NB15}~\cite{moustafa2015unsw} & 2015 & 447,915 & - & 10 & 49 & No\\ 
\hline
 \textbf{CICIDS2017}~\cite{sharafaldin2018toward} & 2017 & 2,830,743 & - & 8  & 84 & No\\ 
\hline
CSE-CIC-IDS2018~\cite{sharafaldin2018toward} & 2018 & 16,136,255 & - &  12 & 76 & No\\ 
\hline
\textbf{CICDDoS2019}~\cite{sharafaldin2019developing} & 2019 & 50,063,112 & -  & 11 & 84 & No\\  
\hline
Car Hacking~\cite{qvr7-n418-21} & 2020 & 8,694,507 & - & 5 & 12 & No\\ 
\hline
 FLNET2023~\cite{kumar2023flnet2023} & 2023 & 6,807,107 & - & 11 & 84 & Yes\\  
\hline
 \textbf{CICIoT2023}~\cite{neto2023ciciot2023} & 2023 & 46,556,613 & - & 25 & 39 & No\\  
\hline
CIC IoV~\cite{neto2024ciciov2024} & 2024 & 1,408,219 & - & 6 & 12 & No \\  
\hline
X-CANIDS~\cite{JeongLLK24X-CANIDS} & 2024 & 286,293 & - & 5 & 689 & No \\
\hline
% \textcolor{red}{FlowTrace-IDB} ~\cite{} & 2026 &  & - &  &  & No \\
% \hline
\end{tabular}
\caption{Representative IDS datasets. Datasets in bold are widely used benchmark datasets in recent IDS research.}
\label{tab:NIDS_dataset}
\end{table*}
%=============================

Benchmark datasets are essential for developing, evaluating, and comparing IDS models because they provide shared traffic records, attack labels, and feature representations for reproducible experiments~\cite{kumar2023flnet2023,sharafaldin2019developing,sharafaldin2018toward,moustafa2015unsw}. In IDS research, dataset quality directly affects the reliability of model evaluation. A dataset should contain representative benign and malicious traffic, diverse attack types, clear labeling rules, and sufficient feature information for downstream detection tasks. However, constructing realistic IDS datasets is difficult because real network traffic may contain sensitive user information, organization-specific configurations, and security-critical infrastructure details.

Several studies have reviewed the limitations and design considerations of IDS datasets. Ring \textit{et al.}~\cite{ring2019survey} analyzed network-based IDS datasets from the perspective of attack scenarios and discussed the relationships among different datasets. Their work also provided recommendations for dataset selection and future dataset construction. Khraisat \textit{et al.}~\cite{khraisat2019survey} reviewed IDS datasets together with detection techniques, data collection methods, evaluation practices, and dataset limitations. These studies show that dataset selection is not only an experimental detail but also a key factor that determines whether IDS results can generalize to realistic deployment environments.

Table~\ref{tab:NIDS_dataset} summarizes commonly used IDS datasets and recent domain-specific datasets. The listed datasets cover traditional network intrusion detection, distributed denial-of-service detection, IoT traffic, in-vehicle network security, and FL-oriented IDS evaluation. The bold datasets, including UNSW-NB15, CICIDS2017, CICDDoS2019, and CICIoT2023, are emphasized because they are widely used in recent IDS studies and provide relatively large-scale traffic records with multiple attack categories. In contrast, datasets such as Car Hacking, CIC IoV, and X-CANIDS are more domain-specific and are mainly designed for vehicle or controller area network security. FLNET2023 is particularly relevant to FL-based IDS because it explicitly supports federated evaluation, while most existing IDS datasets are centrally collected and do not naturally reflect client-level data distribution.

Although these datasets have supported substantial IDS research, several limitations remain. Many benchmark datasets are collected in controlled environments and may not fully capture real network topology, user behavior, temporal dynamics, or evolving attack strategies. In addition, most datasets are designed for centralized learning and provide limited support for studying non-independent and identically distributed (non-IID) clients in FL-based IDS. These limitations motivate the development of more realistic, privacy-aware, and federated IDS benchmarks.

\section{Generative models}\label{sec:generative_models}

%------------------------------------------------------------------------------------------------------
\begin{figure}[t]
    \centering
    \begin{forest}
      for tree={
        grow=east,         % Tree grows to the right (east)
        parent anchor=east,% Parent node connects to the east side (right)
        child anchor=west, % Child node connects to the west side (left)
        edge={-},          % Draw a simple line between nodes
        l sep+=10pt,       % Decrease level separation for a more compact layout
        s sep+=15pt,       % Decrease sibling separation for a more compact layout
        anchor=base west,  % Align nodes to the left side
        tier/.wrap pgfmath arg={tier#1}{level()}, % Tier numbering
        font=\small,    % Use a sans-serif font for better readability
        scale=1.0         % Scale down the entire diagram
      }
      [Generative AI with IDS
        [Large Language Model (LLM)\\~\ref{subsec:LLM_in_IDS}
          % [Generative AI Related Surveys]
          % [Federated Learning]
          % [Explainable IDS]
          % [Adversarial Training]
          % [Application in IDS]
          % [Tabular Data Generation]
        ]
        [Generative Diffusion\\~\ref{subsec:Diffusion_IDS}
          % [Additional Relevant Papers]
          % [Generative AI Related Surveys]
          % [Federated Learning]
          % [Adversarial Detection]
          % [Adversarial Training]
          % [Tabular Data Generation]
        ]
        [Generative Adversarial \\
        Networks (GAN)~\ref{subsec:GAN_IDS}
          % [General Diffusion]
          % [Addressing Class Imbalance]
          % [Application in IoT]
          % [Generative AI Related Surveys]
          % [Federated Learning]
          % [Intrusion Detection]
          % [Tabular Data Generation]
        ]
        [Variational Autoencoder\\(VAE)~\ref{subsec:VAE_IDS}
          % [Data Augmentation]
          % [Data Generation]
          % [Anomaly Detection]
        ]
      ]
    \end{forest}
    \caption{Generative AI with intrusion detection system}
    \label{fig:AI_IDS}
\end{figure}
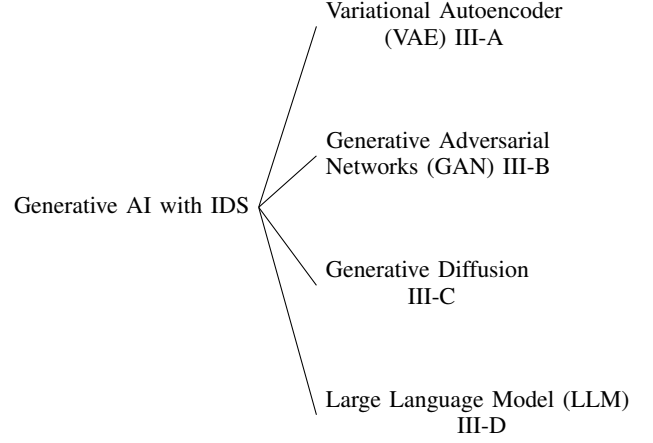
%------------------------------------------------------------------------------------------------------

Generative AI refers to a class of models that learn the underlying structure or probability distribution of observed data and use the learned representation to generate new samples. In the IDS domain, generative models are mainly used to address data-related limitations, including limited attack samples, class imbalance, missing values, and insufficient traffic diversity. They are also used for anomaly detection, adversarial traffic generation, and explanation of IDS outputs. As shown in Figure~\ref{fig:AI_IDS}, this survey organizes generative AI techniques for IDS into four representative model families: Variational Autoencoders (VAEs), Generative Adversarial Networks (GANs), diffusion models, and Large Language Models (LLMs).

Section~\ref{subsec:VAE_IDS} first reviews VAE-based IDS studies. VAEs encode input data into a latent distribution and reconstruct or generate samples from this latent space, making them useful for anomaly detection, data generation, and data augmentation. Section~\ref{subsec:GAN_IDS} then discusses GAN-based IDS methods. GANs use a generator and a discriminator in an adversarial training process and have been widely applied to synthetic traffic generation, class-imbalance mitigation, and adversarial attack generation. Section~\ref{subsec:Diffusion_IDS} reviews diffusion-based methods, which generate data through a denoising process and have recently been extended from image and text generation to tabular data generation. Section~\ref{subsec:LLM_in_IDS} discusses LLM-based IDS applications, including traffic-log analysis, tabular data generation, attack classification, and IDS alert explanation.

Network traffic data used in IDS can appear in different formats, including flow-level tabular features, packet-level byte sequences, traffic logs, and raw packet capture (PCAP) files. Therefore, different generative model families can be applied depending on how the traffic is represented. Tabular generative models are suitable for flow-based IDS datasets, sequence and language models can be used for logs or byte-level representations, and image-based generative models can be applied when traffic records are transformed into image-like formats. The following subsections summarize how these generative models have been used in IDS and what technical challenges remain for each model family.
%------------------------------------------------------------------------------------------------------
\subsection{Variational Autoencoder in IDS}\label{subsec:VAE_IDS}

Variational Autoencoders (VAEs) are generative neural networks introduced by Kingma and Welling~\cite{DBLP:journals/corr/KingmaW13}. A VAE encodes input data into a latent space and reconstructs the input from the learned latent representation. Different from a conventional autoencoder, a VAE models the latent representation as a probability distribution rather than a fixed vector. This probabilistic design allows the model to sample from the latent space and generate new data points that follow the learned data distribution. Therefore, VAEs are suitable for both representation learning and synthetic data generation.

In IDS research, VAEs are mainly used for anomaly detection, data generation, and data augmentation. For anomaly detection, a VAE learns the distribution of normal or known traffic and identifies samples that are difficult to reconstruct. For data generation and augmentation, a VAE can generate synthetic network traffic samples to improve data diversity, mitigate class imbalance, or support downstream IDS training. This subsection reviews VAE-based IDS studies according to these major application objectives.

\subsubsection{Anomaly Detection}
\begin{figure}[t]
    \centering
    \begin{forest}
      for tree={
        grow=east,         % Tree grows to the right (east)
        parent anchor=east,% Parent node connects to the east side (right)
        child anchor=west, % Child node connects to the west side (left)
        edge={-},          % Draw a simple line between nodes
        l sep+=0pt,       % Decrease level separation for a more compact layout
        s sep+=5pt,       % Decrease sibling separation for a more compact layout
        anchor=base west,  % Align nodes to the left sidetier/.wrap pgfmath arg={tier#1}{level()}, % Tier numbering
        font=\scriptsize,    % Use a sans-serif font for better readability
        scale=1.0,        % Scale down the entire diagram
        align=center       % Center-align the text within nodes
      }
          [VAE Anomaly Detection
            [Evaluation
                [Case Studies~\cite{DBLP:conf/kse/NguyenNHS22, DBLP:conf/csr2/HannanGS21}]
                [Performance Metrics~\cite{DBLP:conf/dsc/XieLJJHH19, DBLP:conf/nsyss/AzminI20}]
            ]
            [Applications
                [Critical Infrastructure~\cite{zheng2019anomaly, DBLP:conf/eusipco/TakiddinIZS20}]
                [Insider Threat Detection~\cite{DBLP:conf/csr2/PantelidisBSK21}]
                [Network Traffic Analysis~\cite{DBLP:conf/cns/NguyenLDLC19, DBLP:journals/sensors/KhanamAIJ22}]
            ]
            [Techniques
                [Latent Space Analysis~\cite{DBLP:journals/tjs/ChoiKLK19, DBLP:conf/esorics/OsadaON17}]
                [Reconstruction Error~\cite{DBLP:journals/iotj/XuLYS21, DBLP:journals/access/ZavrakI20}]
            ]
          ]
    \end{forest}
    \caption{Variational autoencoder anomaly detection in intrusion detection system 
    % \textcolor{red}{(title not correct)}
    }
    \label{fig:VAE_Anomaly_Detection}
\end{figure}
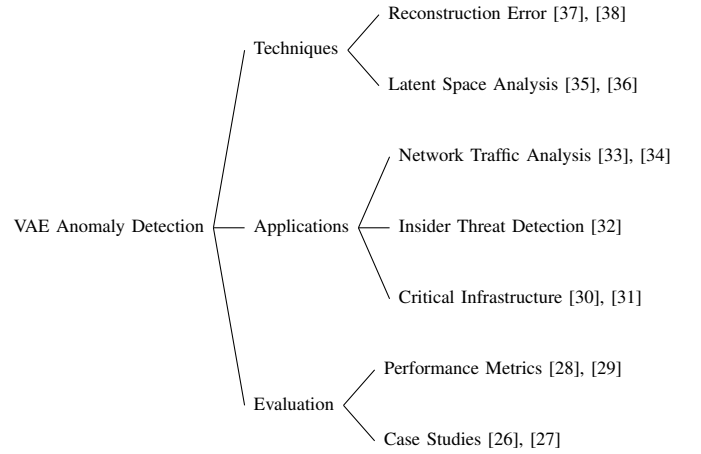

VAE-based anomaly detection relies on the assumption that normal traffic can be reconstructed more accurately than abnormal or unseen traffic. After training, the VAE reconstructs an input sample from its latent representation, and the difference between the original input and the reconstructed output is measured as the reconstruction error. A large reconstruction error indicates that the input does not follow the learned traffic distribution and may correspond to malicious or abnormal behavior. As shown in Figure~\ref{fig:VAE_Anomaly_Detection}, this mechanism has been widely used in anomaly-based IDS.

Several studies have applied VAEs to network-flow anomaly detection. Zavrak and Iskefiyeli~\cite{DBLP:journals/access/ZavrakI20} used VAE-based reconstruction error to detect anomalies from network flow features. Xu \textit{et al.}~\cite{DBLP:journals/iotj/XuLYS21} proposed a Log-Cosh Conditional VAE, where the log-cosh loss improves robustness to outliers during reconstruction. These studies show that reconstruction-based VAE models can provide an effective unsupervised or semi-supervised mechanism for detecting abnormal traffic patterns.

VAEs have also been extended to specific IDS scenarios. In IoT environments, Khanam \textit{et al.}~\cite{DBLP:journals/sensors/KhanamAIJ22} incorporated focal loss into a VAE-based model to improve the detection of abnormal traffic under imbalanced conditions. Nguyen \textit{et al.}~\cite{DBLP:conf/cns/NguyenLDLC19} proposed the Gradient-based Explainable Variational Autoencoder (GEE), which combines anomaly detection with gradient-based explanations to help interpret detected anomalies. For insider threat detection, Pantelidis \textit{et al.}~\cite{DBLP:conf/csr2/PantelidisBSK21} used autoencoder-based models, including VAEs, to identify abnormal user behavior within an organization. In critical infrastructure, VAE-based methods have been applied to power systems and Advanced Metering Infrastructure (AMI) networks to detect data corruption and stealth cyber-attacks~\cite{zheng2019anomaly,DBLP:conf/eusipco/TakiddinIZS20}.

Evaluation studies further show that VAE-based anomaly detection can be improved through model design and loss-function selection. Conditional VAEs have been evaluated for network intrusion detection and shown to be effective when the model is designed to capture class- or condition-specific traffic patterns~\cite{DBLP:conf/dsc/XieLJJHH19,DBLP:conf/nsyss/AzminI20}. Other studies have explored latent-space clustering and inferential autoencoder designs to improve anomaly separation and adapt to changing network conditions~\cite{DBLP:conf/kse/NguyenNHS22,DBLP:conf/csr2/HannanGS21}. Overall, VAE-based anomaly detection is useful for IDS because it can model normal traffic without requiring extensive labeled attack samples, but its performance depends on the quality of the learned latent representation and the threshold used to distinguish normal and abnormal reconstruction behavior.

\subsubsection{Data Generation}
\begin{figure}[t]
    \centering
    \begin{forest}
      for tree={
        grow=east,         % Tree grows to the right (east)
        parent anchor=east,% Parent node connects to the east side (right)
        child anchor=west, % Child node connects to the west side (left)
        edge={-},          % Draw a simple line between nodes
        l sep+=0pt,       % Decrease level separation for a more compact layout
        s sep+=0pt,       % Decrease sibling separation for a more compact layout
        anchor=base west,  % Align nodes to the left sidetier/.wrap pgfmath arg={tier#1}{level()}, % Tier numbering
        font=\scriptsize,    % Use a sans-serif font for better readability
        scale=1.0,        % Scale down the entire diagram
        align=center       % Center-align the text within nodes
      }
          [Data Generation
            [Evaluation
                [Effectiveness in Training IDS Models\\~\cite{DBLP:conf/ictinnovations/GjorgievG20, DBLP:journals/sj/GuhaCS23}]
            ]
            [Use Cases
                [Balancing Imbalanced Datasets\\~\cite{DBLP:journals/tjs/ChuangH23,yang2020network}]
                [Simulation of Network Traffic\\~\cite{DBLP:journals/iotj/DinhNHNBD24, DBLP:journals/sensors/YangZWY19}]
            ]
            [Synthetic Data\\Creation
                [Enhancing Training Dataset\\~\cite{DBLP:journals/sensors/RenFHCC23, DBLP:journals/access/LinLHNLL22}]
                [Methods for Generating \\synthetic data~\cite{DBLP:journals/kais/MartinCS19, DBLP:journals/sensors/MartinCSL17}]
            ]
          ]
    \end{forest}
    \caption{Variational autoencoder in intrusion detection system}
    \label{fig:VAE_Data_Generation}
\end{figure}

VAEs can be used to generate synthetic IDS data by sampling from the learned latent distribution and decoding the sampled representations into new traffic records. This capability is useful when real network traffic is limited, sensitive, or imbalanced. In IDS, VAE-based data generation is mainly used to enrich training datasets, simulate network traffic, and generate samples for underrepresented attack classes. Ren \textit{et al.}~\cite{DBLP:journals/sensors/RenFHCC23} and Lin \textit{et al.}~\cite{DBLP:journals/access/LinLHNLL22} showed that VAE-generated data can improve IDS training by increasing data diversity. Martin \textit{et al.}~\cite{DBLP:journals/kais/MartinCS19,DBLP:journals/sensors/MartinCSL17} further developed VAE-based generative models for intrusion detection and demonstrated their ability to produce synthetic traffic samples for downstream detection tasks.

Another application of VAE-based data generation is traffic simulation. In this setting, the generated samples are used to represent different network behaviors, including benign activities and malicious attacks, under controlled experimental conditions. Dinh \textit{et al.}~\cite{DBLP:journals/iotj/DinhNHNBD24} and Yang \textit{et al.}~\cite{DBLP:journals/sensors/YangZWY19} used VAE-based models to support IDS training by generating or reconstructing traffic patterns that improve the representation of network behaviors. This is useful for evaluating IDS models when collecting sufficient real-world attack traffic is difficult.

VAEs have also been used to address class imbalance by generating synthetic samples for minority attack classes. Class imbalance is common in IDS datasets because benign traffic and frequent attack types often dominate the data, while rare attacks have limited training samples. Chuang and Huang~\cite{DBLP:journals/tjs/ChuangH23} and Yang \textit{et al.}~\cite{yang2020network} showed that VAE-based balancing strategies can improve IDS performance on underrepresented classes. Evaluation studies also indicate that the usefulness of generated data should be measured through downstream IDS performance, such as detection accuracy, recall, and robustness, rather than only by visual or statistical similarity~\cite{DBLP:conf/ictinnovations/GjorgievG20,DBLP:journals/sj/GuhaCS23}.

\subsubsection{Data Augmentation}
\begin{figure}[t]
    \centering
    \begin{forest}
      for tree={
        grow=east,         % Tree grows to the right (east)
        parent anchor=east,% Parent node connects to the east side (right)
        child anchor=west, % Child node connects to the west side (left)
        edge={-},          % Draw a simple line between nodes
        l sep+=0pt,       % Decrease level separation for a more compact layout
        s sep+=5pt,       % Decrease sibling separation for a more compact layout
        anchor=base west,  % Align nodes to the left sidetier/.wrap pgfmath arg={tier#1}{level()}, % Tier numbering
        font=\scriptsize,    % Use a sans-serif font for better readability
        scale=1.0,        % Scale down the entire diagram
        align=center       % Center-align the text within nodes
      }
          [Data Augmentation
            [Evaluation
                [Comparison with Other\\ Augmentation Methods~\cite{10248439, DBLP:conf/icait/ChuangW19}]
                [Impact on Model Performance~\cite{10582127, DBLP:journals/tr/LiuASZ22}]
            ]
            [Applications
                [Improving Model Robustness~\cite{DBLP:conf/icc/YaegashiTN22, DBLP:journals/tifs/YangCCJT21}]
            ]
            [Techniques
                [Feature Augmentation~\cite{DBLP:conf/globecom/SabeelSEE21}]
                [Oversampling~\cite{mohamed2023comparative, DBLP:journals/sensors/YangZWY19}]
            ]
          ]
    \end{forest}
    \caption{Variational autoencoder in intrusion detection system}
    \label{fig:VAE_Data_Augmentation}
\end{figure}

VAE-based data augmentation aims to expand IDS training data by generating additional samples or feature variations from the learned latent distribution. This is different from general data generation because the main objective is not only to create synthetic traffic, but also to improve downstream IDS training. In particular, VAE-based augmentation is useful when the original dataset is limited, imbalanced, or insufficiently diverse.

One common strategy is feature-level augmentation, where a VAE learns latent representations of traffic records and generates variations that enrich the original feature space. This strategy can help the IDS model observe broader traffic patterns during training and reduce overfitting to limited samples. For example, Sabeel \textit{et al.}~\cite{DBLP:conf/globecom/SabeelSEE21} applied VAE-based augmentation to improve the detection of atypical attack flows, where rare or unusual attack behaviors are difficult to learn from the original data alone.

Another important strategy is oversampling, where VAEs generate synthetic samples for underrepresented classes. As noted above, when minority attack classes have too few samples, the detector may become biased toward majority classes and fail to recognize rare attacks. VAE-based oversampling mitigates this issue by increasing the number of minority-class samples while preserving the learned data distribution~\cite{mohamed2023comparative,DBLP:journals/sensors/YangZWY19}.

Several studies have evaluated the impact of VAE-based augmentation on IDS performance. Compared with traditional augmentation or resampling methods, VAE-based approaches can generate samples that better reflect the underlying feature relationships in network traffic~\cite{10248439,DBLP:conf/icait/ChuangW19}. Empirical results also show that VAE-augmented data can improve detection accuracy, robustness, and class-level performance, especially under imbalanced or limited-data settings~\cite{10582127,DBLP:journals/tr/LiuASZ22}.

VAE-based augmentation has also been explored for improving IDS robustness. By exposing the detector to reconstructed, perturbed, or boundary-related traffic samples, the augmented training data can help the model become less sensitive to small variations in traffic patterns. For example, VAE-based methods have been used in two-stage detection and known/unknown intrusion detection settings to improve the model's ability to handle abnormal or unseen traffic~\cite{DBLP:conf/icc/YaegashiTN22,DBLP:journals/tifs/YangCCJT21}. Overall, VAE-based data augmentation supports IDS by improving training data diversity, but its effectiveness still depends on the quality of generated samples and their consistency with realistic network behavior.

%------------------------------------------------------------------------------------------------------

\subsection{GAN in IDS}\label{subsec:GAN_IDS}

%------------------------------------------------------------------------------------------------------

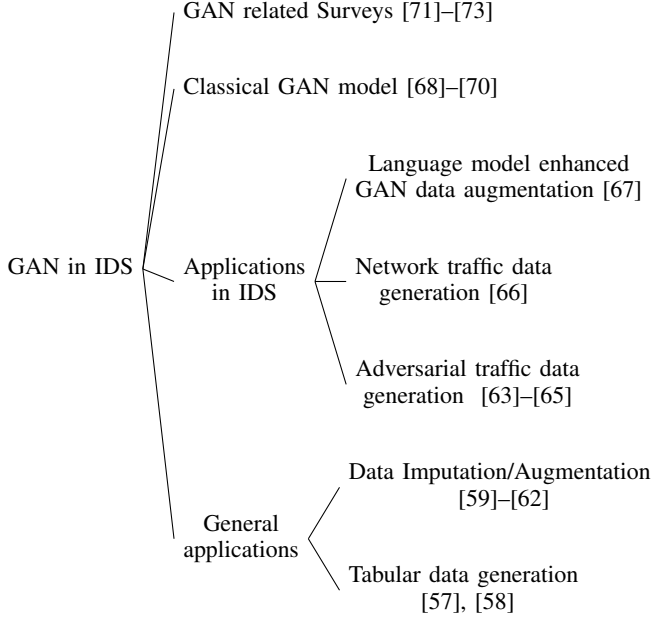
\begin{figure}[hbtp]
    \centering
    \begin{forest}
      for tree={
        grow=east,         % Tree grows to the right (east)
        parent anchor=east,% Parent node connects to the east side (right)
        child anchor=west, % Child node connects to the west side (left)
        edge={-},          % Draw a simple line between nodes
        l sep+=0pt,       % Decrease level separation for a more compact layout
        s sep+=5pt,       % Decrease sibling separation for a more compact layout
        anchor=base west,  % Align nodes to the left sidetier/.wrap pgfmath arg={tier#1}{level()}, % Tier numbering
        font=\small,    % Use a sans-serif font for better readability
        scale=1.0,        % Scale down the entire diagram
        align=center       % Center-align the text within nodes
      }
        [GAN in IDS
            [General \\applications
                [Tabular data generation\\~\cite{DBLP:journals/pvldb/ParkMGJPK18,ctgan}]
                [Data Imputation/Augmentation\\~\cite{DBLP:conf/icaiic/KimTS20, DBLP:journals/access/ShahbazianG23,park2022enhanced,huang2020igan}]
            ]
            [Applications \\in IDS
                [Adversarial traffic data \\generation ~\cite{de2023unsupervised,DBLP:journals/sensors/AldhaheriA23, shu2020generative}]
                [Network traffic data \\generation~\cite{zhao2024enhancing}]   
                [Language model enhanced\\ GAN data augmentation~\cite{DBLP:journals/ppna/LiSMWDM24}]
            ]
            [Classical GAN model~\cite{goodfellow2014generative,gulrajani2017improved,DBLP:journals/corr/ArjovskyCB17}
            ]
            [GAN related Surveys~\cite{DBLP:journals/csur/KammounSTOA23,DBLP:conf/icmv/AissaMZ19,DBLP:journals/pr/RosaP21}]
          ]
        ]
    \end{forest}
    \caption{Generative Adversarial Nets (GAN).}
    \label{fig:GAN_IDS}
\end{figure}

%------------------------------------------------------------------------------------------------------
 
Generative Adversarial Networks (GANs) were introduced by Goodfellow \textit{et al.}~\cite{goodfellow2014generative}. A GAN contains two neural networks: a generator and a discriminator. The generator learns to produce synthetic samples, while the discriminator learns to distinguish generated samples from real samples. Through this adversarial training process, the generator gradually improves its ability to produce data that resemble the original distribution. Later studies improved the stability and quality of GAN training, including Wasserstein GAN and improved Wasserstein GAN~\cite{DBLP:journals/corr/ArjovskyCB17,gulrajani2017improved}. As summarized in Figure~\ref{fig:GAN_IDS}, GANs have been applied to IDS mainly for tabular data generation, data augmentation, data imputation, network traffic generation, and adversarial traffic generation.

\subsubsection{Tabular GAN} \label{subsubsec:tab_GAN}
Many IDS datasets are represented as tabular data, where each row corresponds to a traffic record and each column corresponds to a feature such as duration, packet count, byte count, protocol type, or flow statistic. Therefore, tabular GAN models are relevant to IDS because they can generate synthetic flow-level records for training or evaluation. Park \textit{et al.}~\cite{DBLP:journals/pvldb/ParkMGJPK18} introduced Table-GAN for synthetic table generation, aiming to reduce the risk of exposing real data during data sharing. Xu \textit{et al.}~\cite{ctgan} proposed CTGAN and TVAE to model complex tabular distributions, including mixed continuous and discrete features. These tabular generation methods provide the methodological basis for applying GANs to flow-based IDS datasets.

Conditional GANs have also been used to generate class-specific IDS samples. Li \textit{et al.}~\cite{DBLP:journals/ppna/LiSMWDM24} proposed a BERT-enhanced Conditional GAN for multi-class intrusion detection. In this framework, the CGAN generates additional samples for minority attack classes, while BERT is embedded in the discriminator to strengthen the dependency between input features and output labels. The method addresses class imbalance and improves multi-class IDS performance on several datasets, including CSE-CIC-IDS2018, NF-ToN-IoT-V2, and NF-UNSW-NB15-v2.

\subsubsection{IDS Data Generation} \label{subsubsec:GAN_data_generation}
In IDS, GAN-based data generation is mainly used for three purposes. The first purpose is data augmentation, where GANs generate additional samples to enrich the training set and improve downstream detection performance. This is useful when attack samples are limited or when minority attack classes are underrepresented. Park \textit{et al.}~\cite{park2022enhanced} and Huang and Lei~\cite{huang2020igan} applied GAN-based augmentation to address class imbalance in IDS and improve detection performance. Related surveys also show that GANs have been widely studied for imbalance learning and synthetic data generation in broader machine learning settings~\cite{DBLP:journals/jbd/Sauber-ColeK22,DBLP:journals/jbd/SampathMMG21}.

The second purpose is data imputation, where GANs estimate missing or incomplete feature values while preserving the structure of the original data. Missing values can occur because of packet loss, incomplete collection, preprocessing errors, or unavailable traffic attributes. GAN-based imputation methods have been reviewed in~\cite{DBLP:conf/icaiic/KimTS20,DBLP:journals/access/ShahbazianG23}, and they are relevant to IDS because incomplete traffic records can reduce the reliability of model training and evaluation.

The third purpose is adversarial traffic generation. In this setting, GANs generate malicious traffic records that preserve attack functionality while appearing similar to benign or normal traffic. Such generated samples can be used offensively to bypass IDS models or defensively to evaluate and improve IDS robustness. De Araujo-Filho \textit{et al.}~\cite{de2023unsupervised}, Aldhaheri and Alhuzali~\cite{DBLP:journals/sensors/AldhaheriA23}, and Shu \textit{et al.}~\cite{shu2020generative} studied GAN-based adversarial generation against IDS. In addition, Zhao \textit{et al.}~\cite{zhao2024enhancing} investigated GAN-based network traffic generation for improving IDS performance. These studies show that GANs can support IDS model training, but they also introduce security concerns because generated traffic can be used to test or evade detection systems.

\subsection{Diffusion in IDS}\label{subsec:Diffusion_IDS}

\begin{figure}[hbtp]
    \centering
    \begin{forest}
      for tree={
        grow=east,         % Tree grows to the right (east)
        parent anchor=east,% Parent node connects to the east side (right)
        child anchor=west, % Child node connects to the west side (left)
        edge={-},          % Draw a simple line between nodes
        l sep+=0pt,       % Decrease level separation for a more compact layout
        s sep+=5pt,       % Decrease sibling separation for a more compact layout
        anchor=base west,  % Align nodes to the left sidetier/.wrap pgfmath arg={tier#1}{level()}, % Tier numbering
        font=\small,    % Use a sans-serif font for better readability
        scale=1.0,        % Scale down the entire diagram
        align=center       % Center-align the text within nodes
      }
        [Generative Diffusion \\Surveys
            [Text Generation Surveys~\cite{DBLP:journals/peerj-cs/YiCZZZK24,DBLP:conf/ijcai/0009ZZW23}]
            [General Surveys~\cite{cao2024survey}]
          ]
        ]
    \end{forest}
    \caption{Generative diffusion on surveys.}
    \label{fig:diffusion_survey}
\end{figure}
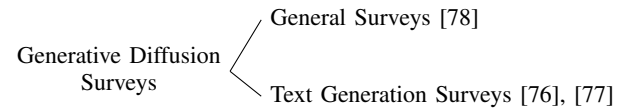

Diffusion models are generative models that learn to generate data through a gradual adding noising and denoising process. Sohl-Dickstein \textit{et al.}~\cite{DBLP:conf/icml/Sohl-DicksteinW15} introduced the diffusion-based generative framework in 2015, where the forward process progressively adds noise to data and the reverse process learns to recover clean samples from noisy inputs. Later, Denoising Diffusion Probabilistic Models (DDPMs) formalized this process as an effective deep generative modeling approach~\cite{ho2020denoising}. Although diffusion models were first widely studied in image generation, they have also been extended to text and tabular data generation~\cite{cao2024survey,DBLP:journals/peerj-cs/YiCZZZK24,DBLP:conf/ijcai/0009ZZW23}. Dhariwal and Nichol~\cite{dhariwal2021diffusion} further showed that diffusion models can achieve strong image generation performance compared with GAN-based methods.

For IDS, diffusion models are relevant because network traffic can be represented in several forms, including tabular flow features, packet sequences, logs, and transformed image-like representations~\cite{villanueva2025netif, yang2022transfer}. Therefore, diffusion models can be used for different IDS-related tasks, including adversarial attack generation, adversarial purification, adversarial training, and synthetic data generation. The following subsections summarize these applications and highlight their relevance to IDS.

\subsubsection{Adversarial Attacks}\label{subsubsec:diff_adv_training}

Diffusion models have been studied in cybersecurity partly because they can generate high-quality synthetic samples that resemble real data. Existing surveys and studies have discussed the role of generative models, including diffusion-related methods, in creating adversarial examples and evaluating security vulnerabilities~\cite{DBLP:journals/cn/NavidanMNSGSW21,DBLP:journals/access/DunmoreJSK23,DBLP:journals/csur/ZhangYTY24,DBLP:conf/uemcom/DuttaGCTB20}. In adversarial attack settings, the generated sample is designed to remain close to a valid input while causing a target ML model to make an incorrect prediction. This idea is relevant to IDS because attackers may modify traffic features while preserving malicious behavior, making the attack harder to detect.

Some diffusion-based adversarial methods combine the denoising process with gradient-based perturbation strategies~\cite{nie2022DiffPure, Chen_2023_ICCV, kumar2026netdiffuser}, such as Projected Gradient Descent (PGD). The purpose is to generate samples that appear realistic but still mislead the target model. Although much of this work has been developed in image domains, the same principle is important for IDS because generated or perturbed traffic can be used to test whether a detector is robust to adaptive attacks.

In response to diffusion-generated fake or adversarial samples, Hooda \textit{et al.}~\cite{DBLP:conf/wacv/HoodaMFFJ024} proposed Disjoint Diffusion Deepfake Detection (D4). D4 is designed to detect fake images generated by diffusion models and to generalize to unseen data distributions and generative techniques. While this work is not specific to IDS, it reflects a broader security problem: as generative models become stronger, detection systems must also be evaluated against generated and previously unseen samples.

\subsubsection{Adversarial Purification}\label{subsubsec:adv_purif}

%------------------------------------------------------------------------------------------------------

\begin{figure}[hbtp]
    \centering
    \begin{forest}
      for tree={
        grow=east,         % Tree grows to the right (east)
        parent anchor=east,% Parent node connects to the east side (right)
        child anchor=west, % Child node connects to the west side (left)
        edge={-},          % Draw a simple line between nodes
        l sep+=0pt,       % Decrease level separation for a more compact layout
        s sep+=5pt,       % Decrease sibling separation for a more compact layout
        anchor=base west,  % Align nodes to the left sidetier/.wrap pgfmath arg={tier#1}{level()}, % Tier numbering
        font=\small,    % Use a sans-serif font for better readability
        scale=1.0,        % Scale down the entire diagram
        align=center       % Center-align the text within nodes
      }
        [Adversarial \\Purification
            [Evaluation~\cite{DBLP:conf/iclr/CarliniTDRSK23,DBLP:conf/iccv/LeeK23, DBLP:journals/corr/abs-2206-10875}
            ]
            [Applications
                [Image domain~\cite{DBLP:journals/corr/abs-2404-14309,DBLP:conf/iwdw/BaoCNE23,DBLP:journals/corr/abs-2205-14969}]
                [3D image domain~\cite{DBLP:journals/corr/abs-2208-09801, DBLP:journals/corr/abs-2403-06698}]
                [Text domain~\cite{DBLP:conf/bigdatasec/Ren0XC24,DBLP:journals/corr/abs-2403-16067}]
                [Audio domain~\cite{DBLP:journals/corr/abs-2310-14270}]
            ]
            [Techniques~\cite{DBLP:conf/iclr/SongKNEK18, DBLP:conf/nips/KangSL23, DBLP:conf/icml/NieGHXVA22}
            ]
          ]
        ]
    \end{forest}
    \caption{Generative diffusion on adversarial purification.}
    \label{fig:diffusion_adversarial_purification}
\end{figure}
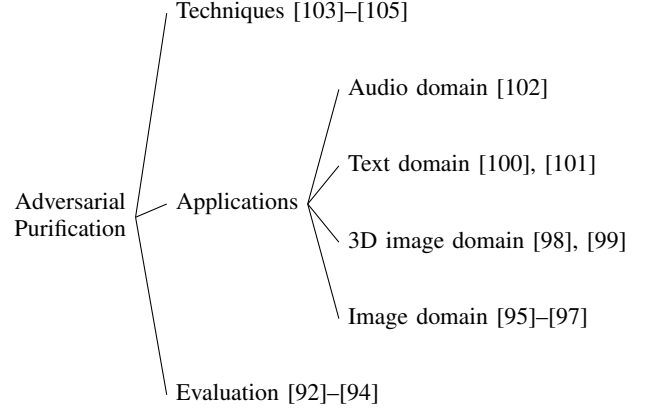

Adversarial purification (AP) is a defense strategy that uses a generative model to remove or reduce adversarial perturbations before classification. The main idea is to map a potentially perturbed input back toward the clean data distribution so that the downstream classifier receives a less corrupted sample. Early work such as PixelDefend used generative modeling for purification~\cite{DBLP:conf/iclr/SongKNEK18}. More recent studies have applied diffusion models to adversarial purification because the denoising process can naturally remove small perturbations from input data~\cite{DBLP:conf/icml/NieGHXVA22,DBLP:journals/corr/abs-2205-14969,DBLP:journals/corr/abs-2206-10875}.

Diffusion-based AP has been evaluated in several domains, including image, 3D point cloud, text, and audio data~\cite{DBLP:journals/corr/abs-2404-14309,DBLP:conf/iwdw/BaoCNE23,DBLP:journals/corr/abs-2208-09801,DBLP:journals/corr/abs-2403-06698,DBLP:conf/bigdatasec/Ren0XC24,DBLP:journals/corr/abs-2403-16067,DBLP:journals/corr/abs-2310-14270}. Carlini \textit{et al.}~\cite{DBLP:conf/iclr/CarliniTDRSK23} further studied the robustness guarantees of diffusion-based purification, while other studies examined its limitations and evaluation reliability~\cite{DBLP:conf/iccv/LeeK23,DBLP:conf/nips/KangSL23}. For IDS, adversarial purification is a potential direction because network traffic may be intentionally perturbed to evade detection. However, applying AP to IDS requires preserving protocol validity and attack semantics, not only removing statistical noise.

\subsubsection{Adversarial Training}\label{subsubsec:diffusion_adv_training}

%------------------------------------------------------------------------------------------------------

\begin{figure}[hbtp]
    \centering
    \begin{forest}
      for tree={
        grow=east,         % Tree grows to the right (east)
        parent anchor=east,% Parent node connects to the east side (right)
        child anchor=west, % Child node connects to the west side (left)
        edge={-},          % Draw a simple line between nodes
        l sep+=0pt,       % Decrease level separation for a more compact layout
        s sep+=5pt,       % Decrease sibling separation for a more compact layout
        anchor=base west,  % Align nodes to the left sidetier/.wrap pgfmath arg={tier#1}{level()}, % Tier numbering
        font=\small,    % Use a sans-serif font for better readability
        scale=1.0,        % Scale down the entire diagram
        align=center       % Center-align the text within nodes
      }
        [Adversarial\\ Training
            [Application\cite{DBLP:journals/corr/abs-2312-04382}
            ]
            [
            Data driven
            % Performance\\ optimization
                [Data size is important\\~\cite{DBLP:conf/nips/SchmidtSTTM18,DBLP:conf/cvpr/Stutz0S19}]
                [Expand dataset size\\~\cite{DBLP:conf/nips/DongDP0020,DBLP:conf/iclr/WangY19,DBLP:conf/aistats/JiangCSDZ21, DBLP:conf/icml/WangPDL0Y23}]
            ]
            [Adversarial Robustness~\cite{DBLP:journals/corr/GoodfellowSS14}]
            ]          ]
        ]
    \end{forest}
    \caption{Generative diffusion on adversarial training.}
    \label{fig:diffusion_adversarial_training}
\end{figure}
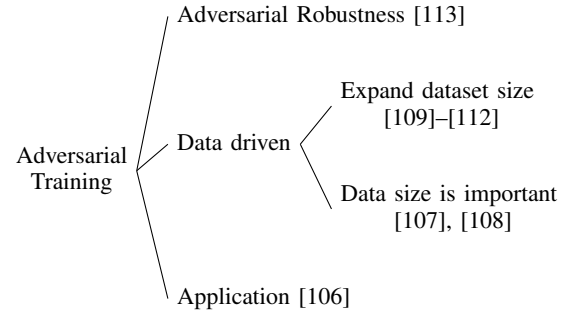

Adversarial training (AT) improves model robustness by training the model with adversarially perturbed examples. Goodfellow \textit{et al.}~\cite{DBLP:journals/corr/GoodfellowSS14} introduced adversarial examples and demonstrated that including such examples during training can improve resistance to adversarial attacks. Later studies showed that robust generalization often requires more training data, because the model must learn stable decision boundaries under both clean and adversarial conditions~\cite{DBLP:conf/nips/SchmidtSTTM18,DBLP:conf/cvpr/Stutz0S19}. This observation motivated the use of external data and generative models to expand training sets for robust learning~\cite{DBLP:conf/nips/DongDP0020,DBLP:conf/iclr/WangY19,DBLP:conf/aistats/JiangCSDZ21}.

Diffusion models can support adversarial training by generating additional training samples that improve data diversity. Wang \textit{et al.}~\cite{DBLP:conf/icml/WangPDL0Y23} used an Elucidating Diffusion Model (EDM) to generate high-quality synthetic image data for adversarial training and showed that generated data can improve robustness without relying only on external real data. Yu \textit{et al.}~\cite{DBLP:journals/corr/abs-2312-04382} proposed the Adversarial Denoising Diffusion Model (ADDM) for unsupervised anomaly detection and showed that it can improve performance compared with DDPM-based anomaly detection methods under reduced sample settings. These studies suggest that diffusion-generated data may be useful for improving IDS robustness, especially when real adversarial or rare attack samples are limited.

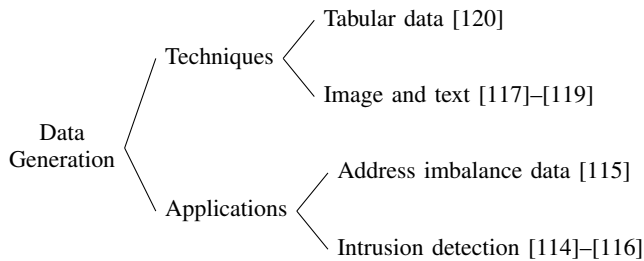
\begin{figure}[hbtp]
    \centering
    \begin{forest}
      for tree={
        grow=east,         % Tree grows to the right (east)
        parent anchor=east,% Parent node connects to the east side (right)
        child anchor=west, % Child node connects to the west side (left)
        edge={-},          % Draw a simple line between nodes
        l sep+=0pt,       % Decrease level separation for a more compact layout
        s sep+=5pt,       % Decrease sibling separation for a more compact layout
        anchor=base west,  % Align nodes to the left sidetier/.wrap pgfmath arg={tier#1}{level()}, % Tier numbering
        font=\small,    % Use a sans-serif font for better readability
        scale=1.0,        % Scale down the entire diagram
        align=center       % Center-align the text within nodes
      }
        [Data\\ Generation
            [Applications
                [Intrusion detection~\cite{DBLP:journals/sensors/TangLLBYY23,wang2023intrusion,lee2023data}]
                [Address imbalance data~\cite{wang2023intrusion}]
            ]
            [Techniques
                [Image and text~\cite{DBLP:conf/nips/HoogeboomNJFW21, DBLP:conf/nips/YeWL22, DBLP:conf/iclr/ChenZH23}]
                [Tabular data~\cite{DBLP:conf/icml/KotelnikovBRB23}]
            ]
        ]
        ]
    \end{forest}
    \caption{Generative diffusion on tabular data generation.}
    \label{fig:Diffusion_tabular}
\end{figure}

%------------------------------------------------------------------------------------------------------
\subsubsection{Data Generation}\label{subsubsec:diff_tabular_gen}
Data generation is the most direct application of diffusion models to IDS~\cite{liu2025synthetic}. Many IDS datasets are represented as tabular flow-level features, where each record describes a network flow using statistical attributes such as packet counts, byte counts, duration, and protocol information. Therefore, tabular diffusion models are particularly relevant to IDS. TabDDPM~\cite{DBLP:conf/icml/KotelnikovBRB23} extended diffusion modeling to tabular data and provided a basis for generating structured records with both numerical and categorical features.

Recent studies have applied diffusion models to intrusion detection and related security tasks~\cite{DBLP:journals/sensors/TangLLBYY23,wang2023intrusion,lee2023data}. In these settings, diffusion-generated samples can be used to increase data diversity, supplement limited attack samples, or mitigate class imbalance. Wang \textit{et al.}~\cite{wang2023intrusion} further showed that diffusion-based generation can help address data imbalance by generating additional samples for underrepresented categories. Compared with GAN-based generation, diffusion models may provide more stable training~\cite{liu2025synthetic}, but their usefulness for IDS still depends on whether the generated traffic preserves realistic feature relationships, protocol constraints, and attack semantics.

\subsection{LLM in IDS}\label{subsec:LLM_in_IDS}

%------------------------------------------------------------------------------------------------------

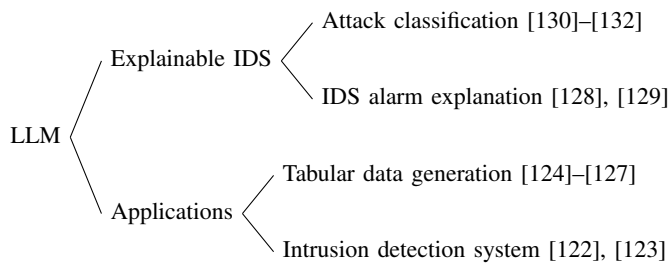
\begin{figure}[hbtp]
    \centering
    \begin{forest}
      for tree={
        grow=east,         % Tree grows to the right (east)
        parent anchor=east,% Parent node connects to the east side (right)
        child anchor=west, % Child node connects to the west side (left)
        edge={-},          % Draw a simple line between nodes
        l sep+=0pt,       % Decrease level separation for a more compact layout
        s sep+=5pt,       % Decrease sibling separation for a more compact layout
        anchor=base west,  % Align nodes to the left sidetier/.wrap pgfmath arg={tier#1}{level()}, % Tier numbering
        font=\small,    % Use a sans-serif font for better readability
        scale=1.0,        % Scale down the entire diagram
        align=center       % Center-align the text within nodes
      }
        [LLM
            [Applications
                [Intrusion detection system~\cite{DBLP:conf/aina/LiraMT24,DBLP:journals/eswa/ManocchioLLKSP24}]
                [Tabular data generation~\cite{kim2024exploring,seedat2023curated,einy2024cost,xu2024llms}]
            ]
            [Explainable IDS
                [IDS alarm explanation~\cite{rjoub2023survey,juttner2023chatids}]
                [Attack classification~\cite{ali2023huntgpt,rao2021zero,DBLP:conf/amcis/ZhongNC23}]
            ]
        ]
        ]
    \end{forest}
    \caption{Large Language Model (LLM).}
    \label{fig:LLM_IDS}
\end{figure}

%------------------------------------------------------------------------------------------------------

Large Language Models (LLMs) are transformer-based models trained to process and generate sequential data. In IDS research, LLMs and related transformer-based language models can be applied when network traffic is represented as logs, packet-byte sequences, flow records converted into textual formats, or structured tabular records. Compared with conventional ML models, LLMs provide two potential advantages for IDS: they can model contextual relationships in sequential traffic representations, and they can generate natural-language explanations for security analysts. However, their use in IDS also introduces practical challenges, including high computation cost, data formatting sensitivity, limited interpretability, and possible hallucination when explanations are generated without sufficient grounding.

\subsubsection{IDS Applications}\label{subsubsec:LLM_IDS_applications}

LLM-based and transformer-based IDS studies mainly focus on using language-model architectures to classify network activities. Lira \textit{et al.}~\cite{DBLP:conf/aina/LiraMT24} proposed BERTIDS, a BERT-based model for network intrusion detection. In this method, network logs are converted into tokenized sequences that can be processed by BERT. The model is then fine-tuned to distinguish normal traffic from different attack categories. Their experiments on NSL-KDD reported an accuracy of 98.01\% and a false positive rate of 1.48\%, showing that transformer-based language models can be adapted to IDS classification tasks.

Manocchio \textit{et al.}~\cite{DBLP:journals/eswa/ManocchioLLKSP24} presented FlowTransformer, a modular framework for transformer-based Network Intrusion Detection Systems (NIDSs). The framework allows different components, including input encoding, transformer architecture, classification head, and evaluation dataset, to be replaced and compared. Their evaluation across public flow-based NIDS datasets showed that the classification head has a substantial effect on detection performance. This result indicates that applying transformers to IDS is not only a matter of selecting a large model; the representation of traffic features and the design of the output classifier are also important.

Although these studies show the potential of transformer-based models for IDS, several limitations remain. First, fine-tuning large models requires substantial computation and memory resources, which may be impractical for resource-constrained security environments. Second, IDS data are often numerical or mixed-type tabular records, while LLMs are originally designed for textual sequences. Therefore, the performance of LLM-based IDS depends strongly on how traffic records are encoded into model-readable inputs. Third, model complexity can make it difficult to interpret why a specific traffic record is classified as malicious, which limits direct use in high-stakes security operations.

\subsubsection{LLM for Explainable IDS}\label{subsubsec:LLM_explain_IDS}
Another important use of LLMs in IDS is explanation generation. IDS alerts often contain technical information such as attack labels, traffic features, source and destination attributes, and model confidence scores. These alerts may be difficult for non-expert users to interpret. LLMs can be used as an explanation layer that converts IDS outputs into natural-language descriptions, summarizes possible causes, and suggests response actions. In this setting, the LLM does not necessarily replace the detector; instead, it helps users understand and act on detection results.

Rjoub \textit{et al.}~\cite{rjoub2023survey} discussed the role of explainable AI in cybersecurity and highlighted the need for human-understandable explanations in security systems. Juttner \textit{et al.}~\cite{juttner2023chatids} proposed ChatIDS, which uses ChatGPT to explain IDS alerts and provide suggestions to non-expert users. This type of approach can improve the usability of IDS outputs, especially when alerts must be interpreted by operators who do not have deep knowledge of network security or ML models.

LLMs have also been combined with traditional ML classifiers and XAI tools. Ali \textit{et al.}~\cite{ali2023huntgpt} introduced HuntGPT, an intrusion detection dashboard that integrates a Random Forest classifier, XAI methods such as SHAP and LIME, and GPT-3.5 Turbo. The classifier detects anomalies, the XAI methods identify important features behind the prediction, and the LLM presents the result in a more understandable form for analysts. Other studies have also explored explainable AI for attack classification and cybersecurity analysis~\cite{rao2021zero,DBLP:conf/amcis/ZhongNC23}. These works show that LLMs are useful for improving the communication between IDS models and human users, but the generated explanations should be grounded in detector outputs and verified evidence to avoid unsupported conclusions.

\subsubsection{LLM Tabular Data Generation} \label{subsubsec:LLM_data_generation}

In 2024, Kim et al.~\cite{kim2024exploring} investigated the effectiveness of using LLM to generate synthetic data that addresses the class imbalance in tabular data. The paper found that using CSV-style prompting (compared to sentence-style in GReaT) can significantly improve the ability of LLM to generate accurate and balanced data, enhancing ML performance for minor classes in imbalanced data.

Not all data generated by LLMs is equally valuable and useful to downstream model performance; some samples may be harmful. Thus, assessment of the generated data is vital for any generative model. In 2024, Seedat et al.~\cite{seedat2023curated} introduced a method called Curated LLM (CLLM), which aims to generate synthetic tabular data in environments where data is scarce ($n < 100$) and to apply a rigorous data curation process to ensure the quality of the generated data. The CLLM first harnesses the prior knowledge embedded in LLMs, using them to generate synthetic datasets based on a small number of real examples. Then, it relies on a curation mechanism that uses metrics like predictive confidence and uncertainty to filter and refine the generated data, improving its utility for downstream ML tasks. The paper used several real-world datasets to demonstrate the superior performance of CLLM over conventional generators (including CTGAN, TVAE, TabDDPM, SMOTE, and GReaT) in the low-data regimes.

In many domains where data privacy is crucial, synthetic data generated from real datasets can be used to avoid exposing real-world data. However, synthetic data can still contain the original dataset pattern or details. Differential privacy (DP), which introduces randomness in the data generation, is a promising approach to reducing the risk of re-identifying individuals. In 2024, Tran et al.~\cite{tran2024differentially} introduced DP-LLMTGen (Differentially Private LLM-based Tabular data Generators), a novel framework designed to generate synthetic tabular data while preserving DP. The framework utilizes a two-stage fine-tuning procedure with a novel loss function specifically designed for tabular data. The first stage focuses on learning the data format using non-sensitive, randomly generated data with original sensitive data. The second stage fine-tunes the LLM with DP mechanisms to ensure that the generated data maintains privacy while accurately capturing the feature distributions and dependencies of the original dataset. Then, synthetic data are generated by sampling from the fine-tuned LLM. The experiment result showed that the proposed DP-LLMTGen framework is able to effectively generate high-fidelity synthetic tabular data while preserving differential privacy.

To address the inefficiency and high computational costs associated with using LLMs for tasks involving tabular data, Einy et al.~\cite{einy2024cost} proposed a selective enrichment approach. Their method aims to use LLMs to enrich tabular data to enhance the performance of classical ML models. LLMs are applied only to specific parts of the data that benefit the most from the additional contextual knowledge that LLMs provide. The result demonstrates that this approach can significantly enhance the performance of ML models on tabular data while maintaining cost-effectiveness.

However, in 2024, Xu et al.~\cite{xu2024llms} demonstrated that LLMs are generally inadequate for tabular data generation when used directly or even after traditional fine-tuning. They suggest that due to their autoregressive nature, LLMs struggle to model the complex conditional dependencies and mixture distributions that exist in real-world tabular data. Further, feature ordering becomes more important when the dataset grows, and incorrect feature order can significantly degrade the quality of the generated data. The authors proposed a novel approach called Permutation-aided Fine-tuning (PAFT). Although the results show that PAFT can reproduce underlying relationships in generated data, there is still a significant gap between the current capabilities of LLMs and the requirements for generating realistic synthetic tabular data.

%------------------%---------
\begin{figure*}[htbp]
     \centering
     \begin{subfigure}[b]{0.49\textwidth}
         \centering
         \includegraphics[width=\textwidth]{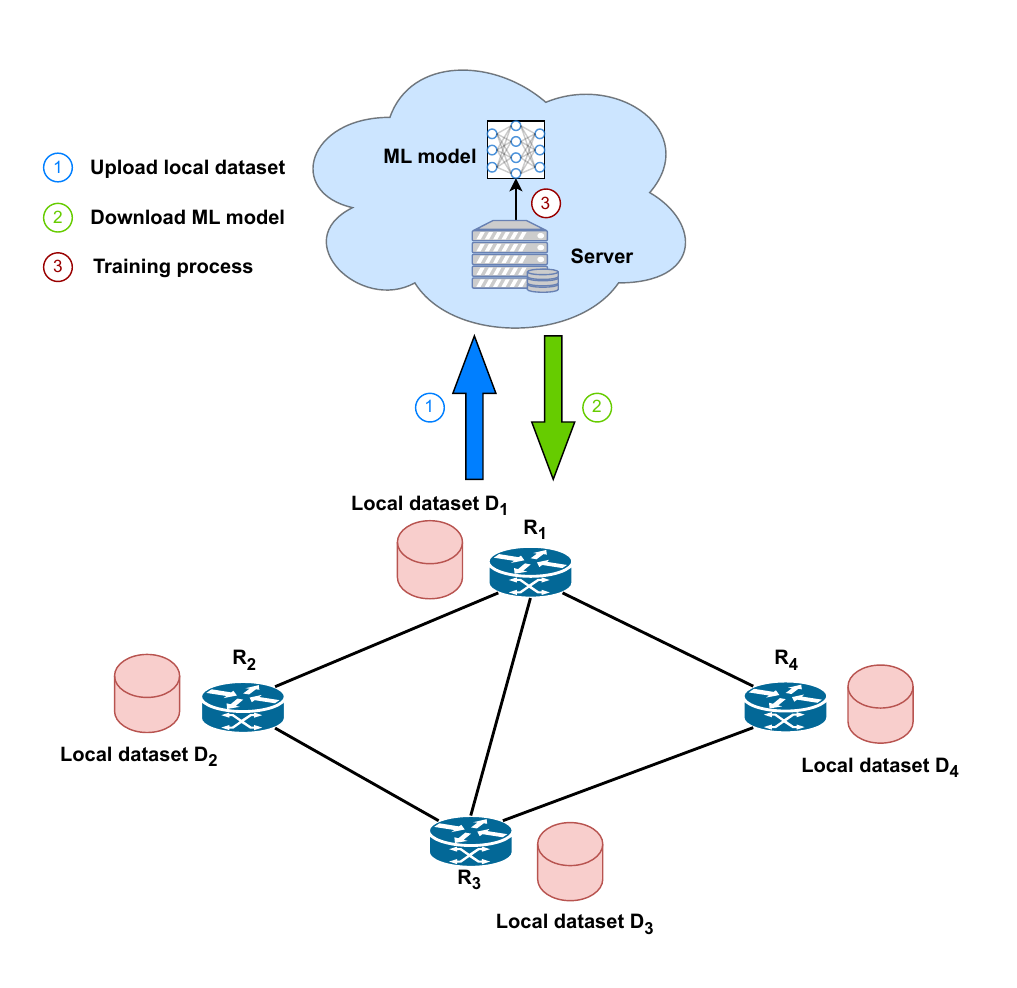}
         \caption{Centralized machine learning}
         \label{fig:Cen_ML}
     \end{subfigure}
     \hfill
     \begin{subfigure}[b]{0.49\textwidth}
         \centering
         \includegraphics[width=\textwidth]{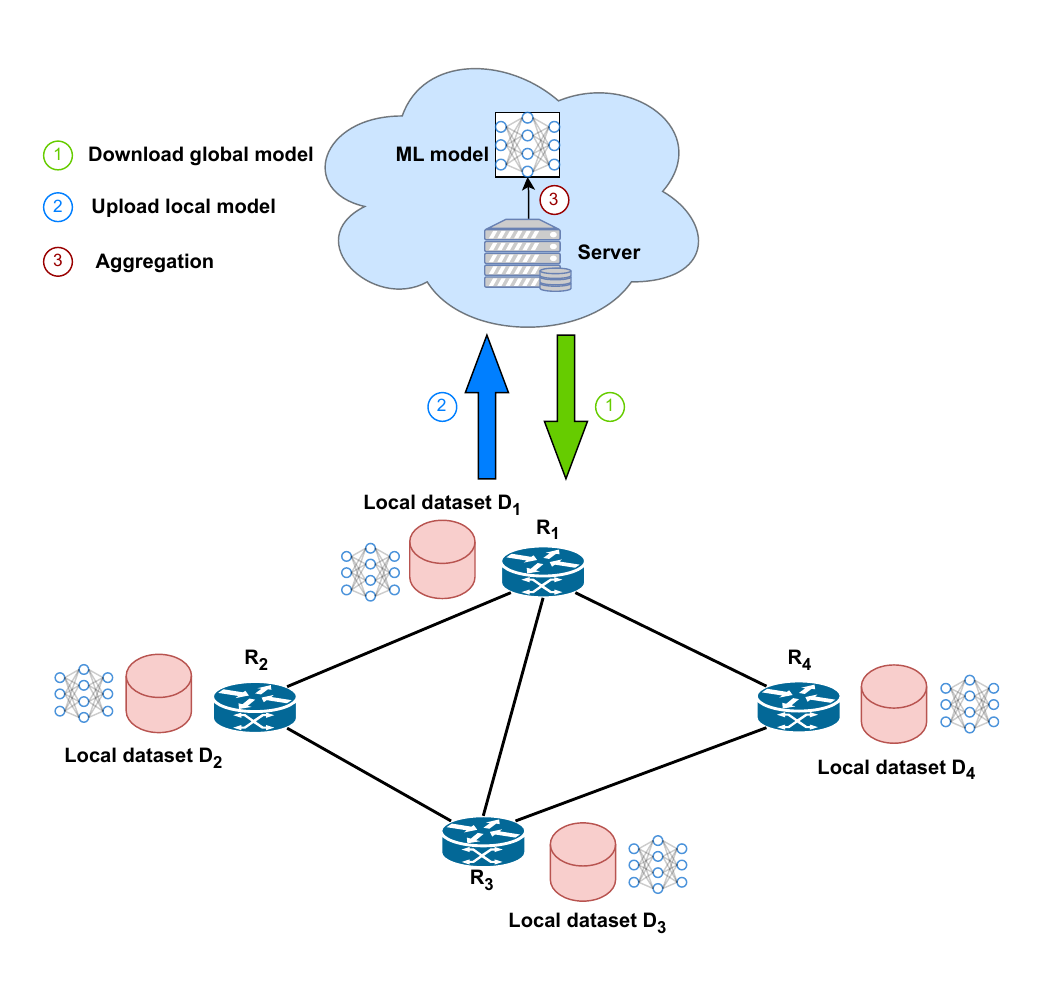}
         \caption{Federated learning}
         \label{fig:FL}
     \end{subfigure}
       % \vspace{-0.17in}
     \caption{Centralized VS Federated Learning}
     \label{fig:cen_vs_FL}
     % \vspace{-0.05in}
\end{figure*}
%=============================

\subsection{Challenges and Valuable Research Directions}\label{subsec:AI_IDS_directions}

Generative AI has been applied to IDS mainly to address data-related limitations. When labeled attack samples are limited, generative models can create additional samples for training. When datasets contain missing or incomplete records, generative models can support data imputation by estimating missing values from learned feature relationships. When datasets are imbalanced, generative models can generate samples for minority attack classes and reduce the bias of IDS models toward majority classes. Although these applications are useful, the use of generated data in IDS also introduces several challenges.

The first challenge is the quality of synthetic data. Generated samples are usually evaluated through their effect on downstream IDS models, such as whether they improve classification accuracy, recall, or robustness. However, synthetic data do not always improve downstream performance. Low-quality samples may introduce noise, distort class boundaries, or cause the detector to overfit artificial patterns. In addition, the performance of generative models is sensitive to design choices and parameter settings, such as latent-space design in VAEs, training stability in GANs, and noise schedules or sampling steps in diffusion models~\cite{kingma2016improved,arjovsky2017towards,DBLP:conf/iclr/CheLJBL17,dhariwal2021diffusion,nichol2021improved}. Therefore, selecting appropriate generative model configurations remains an important problem for IDS applications.

The second challenge is the reliability and realism of synthetic network traffic. Common distribution-level metrics, such as Kullback-Leibler (KL) Divergence~\cite{kullback1951information}, Jensen-Shannon (JS) Divergence~\cite{lin1991divergence}, Wasserstein Distance~\cite{villani2009optimal}, Fréchet Inception Distance (FID)~\cite{heusel2017gans}, Maximum Mean Discrepancy (MMD)~\cite{gretton2012kernel}, Perceptual Path Length (PPL)~\cite{karras2019style}, Energy Distance~\cite{szekely2013energy}, and Precision and Recall for Distributions~\cite{sajjadi2018assessing}, can measure similarity between real and generated data distributions. However, distributional similarity alone is not sufficient for IDS. Synthetic traffic should also preserve protocol constraints, temporal dependencies, attack semantics, and network-topology relationships. For example, generated traffic may be statistically similar to real traffic but still invalid in a real testbed if packet sequences violate protocol behavior or if flows do not match the underlying network topology. Future work should develop IDS-specific evaluation methods that assess both statistical fidelity and network-level validity.

The third challenge is the limited use of Large Language Models (LLMs) for IDS-specific data generation. Existing LLM-based generation methods have shown potential for text and tabular data generation, including prompt-based synthetic data generation~\cite{schmidt2024prompting,desalvo2024no,DBLP:conf/acl/0003XSHTGJ22}. However, IDS data often contain numerical features, protocol-dependent relationships, temporal patterns, and topology-aware constraints, which are difficult for general-purpose LLMs to model directly. At the same time, LLMs provide a potential advantage because they can incorporate textual descriptions, domain knowledge, network configurations, and attack procedures during generation. This makes LLMs a promising direction for topology-aware and knowledge-guided IDS data generation. Future research should investigate how to adapt LLMs to network-security domains, how to ground generated traffic in valid network behavior, and how to evaluate whether LLM-generated samples are useful for IDS training and testing.

Overall, generative AI can support IDS by improving data availability, diversity, and robustness. However, future studies should move beyond simply generating more samples. More attention is needed on sample quality control, IDS-specific realism evaluation, privacy preservation, adversarial misuse, and domain-specific generative models that understand network protocols, attack behaviors, and deployment environments.

\section{Generative AI Embedded Federated Learning based Intrusion Detection System}\label{sec:AI_FL_IDS}

%------------------------------------------------------------------------------------------------------
\begin{figure}[htbp]
    \centering
    \begin{forest}
      for tree={
        grow=east,         % Tree grows to the right (east)
        parent anchor=east,% Parent node connects to the east side (right)
        child anchor=west, % Child node connects to the west side (left)
        edge={-},          % Draw a simple line between nodes
        l sep+=10pt,       % Decrease level separation for a more compact layout
        s sep+=15pt,       % Decrease sibling separation for a more compact layout
        anchor=base west,  % Align nodes to the left side
        tier/.wrap pgfmath arg={tier#1}{level()}, % Tier numbering
        font=\small,    % Use a sans-serif font for better readability
        scale=1.0         % Scale down the entire diagram
      }
      [FL-based IDS \\with generative AI
        [LLM embedded FL 
        ]
        [Diffusion embedded FL~\cite{DBLP:conf/www/JothirajM24}
        ]
        [GAN embedded FL~\cite{DBLP:conf/nss/VyQDP21,DBLP:journals/comcom/TabassumELMG22,DBLP:conf/pakdd/LinSX22,DBLP:journals/sensors/AldhaheriA23,DBLP:conf/hpcc/ZhangZZBZS22}
        ]
        [VAEs embedded FL~\cite{DBLP:journals/mta/NGRN24}
        ]
      ]
    \end{forest}
    \caption{Generative AI with intrusion detection system}
    \label{fig:FL_IDS}
\end{figure}
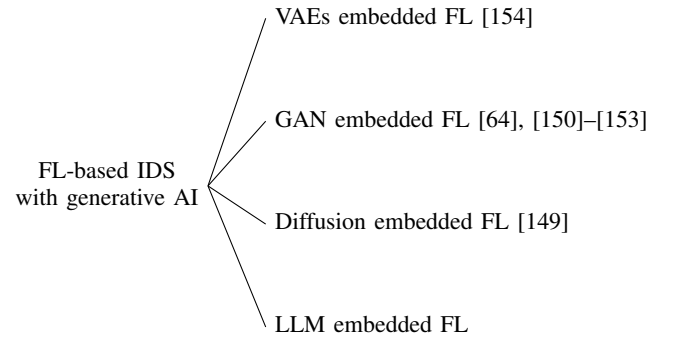
%------------------------------------------------------------------------------------------------------

Federated Learning (FL) is a distributed learning paradigm designed for scenarios where data are generated and stored across many clients. It was originally motivated by applications such as mobile-device intelligence, where user data are distributed across devices and cannot be easily collected at a central server because of privacy and communication constraints~\cite{DBLP:conf/aistats/McMahanMRHA17}. Instead of transferring raw data to the server, FL allows clients to train models locally and share model updates for aggregation. Figure~\ref{fig:cen_vs_FL} illustrates the difference between centralized Machine Learning (ML) and FL. 

In centralized ML, as shown in Figure~\ref{fig:Cen_ML}, clients send their local datasets to a central server. The server then combines the collected data, trains an ML model, and distributes the trained model for future prediction. This approach is simple to implement when data can be centrally collected, but it may expose sensitive information and introduce high communication costs when the local datasets are large.

In FL, as shown in Figure~\ref{fig:FL}, training is performed through repeated collaboration between the server and clients. The server first initializes a global model and sends it to participating clients. Each client trains the model using its local dataset and returns the updated model parameters or gradients to the server. The server then aggregates these local updates to obtain a new global model and sends the updated global model back to clients for the next training round. FedAvg is a representative aggregation method that computes a weighted average of local model updates to construct the global model~\cite{DBLP:conf/aistats/McMahanMRHA17}. By avoiding direct raw-data sharing, FL can reduce privacy risks and raw-data transmission costs, although communication overhead and potential information leakage from model updates remain important concerns.

FL is particularly relevant to IDS because network traffic is naturally distributed across routers, edge devices, organizations, and geographic regions. In a traditional centralized IDS training pipeline, clients or network devices send traffic records to a central server, and the server uses the aggregated data to train a detection model. This process can be expensive for high-volume traffic and may expose sensitive information, such as user behavior, service configurations, or organization-specific security patterns. The problem becomes more significant for high-rate attacks~\cite{munaweera2024federated}, such as denial-of-service (DoS) and distributed denial-of-service (DDoS) attacks, which can generate a large number of traffic records in a short period. FL-based IDS addresses this limitation by allowing each client to train locally while sharing only model updates with the server.

However, FL-based IDS also introduces challenges that are different from centralized IDS. Client data are often non-independent and identically distributed (non-IID)~\cite{liu2023multi} because different clients may observe different traffic volumes, device types, services, and attack categories. Some clients may have limited attack samples, while others may have highly imbalanced traffic distributions. In addition, clients may have different computation and communication capabilities, and the FL process may be vulnerable to poisoning attacks or unreliable updates. These challenges motivate the integration of generative AI with FL-based IDS. Generative models can potentially augment local data, mitigate class imbalance, support privacy-preserving synthetic data generation, improve robustness, and reduce the effect of heterogeneous client distributions.

This section reviews generative AI-embedded FL-based IDS according to the same generative model families discussed in Section~\ref{sec:generative_models}. Section~\ref{subsec:VAE_FL_IDS} discusses VAE-embedded FL-IDS, where autoencoder-based models can support representation learning, anomaly detection, and communication reduction. Section~\ref{subsec:GAN_FL_IDS} reviews GAN-embedded FL-IDS, with a focus on data augmentation, adversarial traffic generation, and class-imbalance mitigation. Section~\ref{subsec:Diffusion_FL_IDS} discusses diffusion-embedded FL-IDS and its potential for synthetic data generation under distributed settings. Finally, Section~\ref{subsec:LLM_FL_IDS} discusses the emerging role of LLMs in FL-based IDS, including their potential use for explanation, knowledge-guided generation, and network-security analysis.

\subsection{Autoencoder-embedded FL-IDS}\label{subsec:VAE_FL_IDS}

Autoencoder-based models, including Autoencoders (AEs) and Variational Autoencoders (VAEs), have been used in FL-based IDS mainly for representation learning, anomaly detection, privacy preservation, and communication reduction. An AE learns to compress input data into a lower-dimensional latent representation and reconstruct the original input from that representation. This design is useful for IDS because network traffic often contains high-dimensional flow features, and the learned latent representation can preserve important traffic patterns while reducing feature size. A VAE extends the AE by modeling the latent representation as a probability distribution, which enables sampling and synthetic data generation from the learned distribution.

In FL-based IDS, autoencoder-based models are useful because clients can learn compact representations of local traffic without directly sharing raw network data. Instead of transmitting full datasets, clients may share model updates or compressed latent representations with the server. This can reduce communication cost and limit direct exposure of sensitive traffic records. In addition, reconstruction error from an AE or VAE can be used for anomaly detection, where traffic samples with large reconstruction errors are treated as potential intrusions~\cite{zha2024skt, DBLP:journals/tifs/YangCCJT21, tang2020zerowall, mirsky2018kitsune}.

Tayeen \textit{et al.}~\cite{tayeen2023cafnet} proposed CAFNET, a compressed autoencoder-based federated framework for network anomaly detection. CAFNET reduces communication overhead by transmitting compact representations between clients and the server rather than transferring raw data or large model components. The reported results show that CAFNET maintains detection performance while reducing communication cost by up to 95\%. NR \textit{et al.}~\cite{DBLP:journals/mta/NGRN24} also used VAE-based federated learning for intrusion detection in Industrial IoT environments, aiming to improve privacy protection and reduce communication overhead.

Overall, AE- and VAE-embedded FL-IDS methods are suitable for settings where network traffic is high-dimensional, privacy-sensitive, and distributed across multiple clients. However, these methods still face several challenges. The compressed latent representation must preserve enough information for accurate intrusion detection, while avoiding unnecessary leakage of sensitive traffic patterns. In addition, when client data are non-IID, the latent spaces learned by different clients may not be fully aligned, which can affect aggregation and global anomaly detection performance.

\subsection{GAN-embedded FL-IDS}\label{subsec:GAN_FL_IDS}

GANs can be integrated with FL-based IDS from two main perspectives. The first perspective treats GAN-generated traffic as a security threat. As discussed in Section~\ref{subsubsec:GAN_data_generation}, GANs can generate adversarial traffic records that resemble real traffic and are designed to bypass IDS models. For example, Lin \textit{et al.}~\cite{DBLP:conf/pakdd/LinSX22} introduced IDSGAN for generating adversarial attack traffic against IDS. Aldhaheri and Alhuzali~\cite{DBLP:journals/sensors/AldhaheriA23} proposed SGAN-IDS, and Zhang \textit{et al.}~\cite{DBLP:conf/hpcc/ZhangZZBZS22} studied poisoning attacks against FL-based network intrusion detection. These studies show that generated malicious traffic can be difficult to detect because it is optimized to remain close to legitimate traffic patterns while misleading the detection model.

In FL-based IDS, this problem becomes more complex because the server does not directly access clients' raw traffic data. The server mainly observes model updates, which makes it harder to inspect whether local training data contain GAN-generated adversarial samples or poisoned records. Therefore, detecting and defending against GAN-based adversarial traffic remains an important challenge for FL-based IDS. Vy \textit{et al.}~\cite{DBLP:conf/nss/VyQDP21} studied poisoning attacks and defense mechanisms in an FL-based IDS framework for Industrial IoT networks, showing that adversarial manipulation must be considered when deploying IDS models in distributed environments.

The second perspective uses GANs as a defensive tool for data augmentation. In FL-based IDS, clients often hold non-IID and imbalanced local datasets because different networks observe different services, devices, traffic volumes, and attack categories. A local GAN can generate additional samples for minority classes and improve the local training distribution without requiring clients to share raw traffic data. Tabassum \textit{et al.}~\cite{DBLP:journals/comcom/TabassumELMG22} proposed FEDGAN-IDS, where GAN-based augmentation is used with FL to address data imbalance while preserving privacy.

Despite this potential, GAN-embedded FL-IDS faces communication and training challenges. A GAN usually contains at least two model components, a generator and a discriminator, which increases training complexity and may introduce additional communication cost if the GAN parameters are shared across clients and the server. Model compression and scaling studies, such as~\cite{li2020gan,hu2022scaling}, show that model size, input size, and performance must be balanced carefully. For FL-based IDS, this means that GANs should be designed or deployed in a way that improves local data quality without offsetting the privacy and communication advantages of FL.

\subsection{Diffusion-embedded FL-IDS}\label{subsec:Diffusion_FL_IDS}

Diffusion models are an emerging direction for generative AI-embedded FL-based IDS. Compared with GANs, diffusion models often provide more stable training because they do not rely on adversarial optimization between a generator and a discriminator. This property is useful in FL, where clients may have heterogeneous and imbalanced local data. However, diffusion models can also require substantial sampling computation, so their communication and computation costs must be evaluated carefully in distributed IDS settings.

Jothiraj and Mashhadi~\cite{DBLP:conf/www/JothirajM24} introduced Phoenix, a federated generative diffusion framework that uses diffusion models to improve training data diversity across clients. Their study compared diffusion-based generation with GAN-based generation in an FL setting and showed that diffusion models can generate high-quality samples while reducing communication cost. Although Phoenix is not limited to IDS, its design is relevant to FL-based IDS because local clients often have limited, imbalanced, and non-IID traffic data.

For FL-based IDS, diffusion models can potentially support local data augmentation, minority-class sample generation, and privacy-preserving synthetic traffic generation. Instead of sharing raw traffic data, clients may use diffusion models to enrich local training data or share generative knowledge with the server. This direction is especially useful when attack samples are rare or unevenly distributed across clients. Nevertheless, diffusion-embedded FL-IDS remains underexplored. Future studies need to examine whether diffusion-generated traffic preserves realistic network behavior, whether the generation process is efficient enough for edge or IoT clients, and how diffusion models can be integrated with FL without increasing communication overhead.

\subsection{LLM-embedded FL-IDS}\label{subsec:LLM_FL_IDS}

LLM-embedded FL-IDS remains an emerging direction with limited direct studies. In this context, an LLM can be integrated into an FL-based IDS pipeline in several ways, including traffic classification, anomaly detection, data augmentation, data imputation, alert explanation, and security knowledge extraction. These capabilities are relevant to FL-based IDS because clients often have limited, imbalanced, and heterogeneous local traffic data. For example, LLM-based data augmentation or imputation may help enrich local datasets without requiring clients to share raw traffic records.

However, applying LLMs directly in FL-based IDS is challenging. The main limitation is the high communication and computation cost of training or fine-tuning large models across distributed clients. Unlike smaller IDS models, LLMs contain a large number of parameters, making full-model transmission impractical for many edge, IoT, or organizational clients. In addition, IDS data are often represented as numerical flow features, packet sequences, or structured logs, while general-purpose LLMs are primarily trained on natural-language text. Therefore, effective input representation and domain adaptation are necessary before LLMs can be reliably used for IDS tasks.

A practical direction is to avoid transmitting full LLMs in the FL process. Instead, future studies may explore parameter-efficient tuning, adapter-based learning, knowledge distillation, or server-side LLM assistance. In these settings, clients may train lightweight local IDS models, while LLMs provide auxiliary functions such as generating domain-informed synthetic samples, explaining alerts, summarizing attack behaviors, or supporting security analysts. Overall, LLM-embedded FL-IDS has potential, but its feasibility depends on reducing communication cost, grounding LLM outputs in valid network behavior, and adapting LLMs to network-security-specific data formats.

\subsection{Challenges and Valuable Research Directions}\label{subsec:FL_AI_IDS_directions}

Generative AI-embedded FL-based IDS introduces both opportunities and risks. On the defensive side, generative models can support local data augmentation, mitigate class imbalance, improve robustness, and reduce the need to share raw traffic data. On the offensive side, the same generative models can be used to create synthetic malicious traffic that resembles benign traffic or bypasses IDS models. This dual-use nature makes generative AI important for both improving FL-based IDS and evaluating its vulnerability under adaptive attacks.

The first research challenge is defending FL-based IDS against generative adversarial traffic. GANs and diffusion models can generate traffic samples that are statistically close to real traffic but intentionally optimized to mislead a detector. In FL-based IDS, this threat is harder to identify because the server usually receives model updates rather than raw local traffic. As a result, adversarial or poisoned local data may influence the global model through aggregation without being directly inspected. Future work should study how to detect generative adversarial traffic in distributed settings, how to distinguish malicious client updates from benign non-IID updates, and how to design robust aggregation methods for FL-based IDS.

The second research direction is using generative AI to address non-independent and identically distributed (non-IID) client data. In FL-based IDS, each client may observe different devices, services, traffic volumes, and attack categories. This heterogeneity can reduce the quality of the global model because local updates are optimized on different data distributions. Generative models can reduce this problem by augmenting local datasets, generating minority-class samples, or improving local data diversity before model training. Existing studies have explored VAE-, GAN-, and diffusion-based methods for handling non-IID data in FL~\cite{yang2023fedvae,morafah2024stable,liu2024slaugfl}. For IDS, this direction is especially relevant because rare attacks may appear only on a small subset of clients.

The third challenge is communication efficiency. Although generative models can improve local training data, transmitting large generative models between the server and clients may increase communication cost. This problem is more significant for GANs and LLMs because they may contain large model components or require expensive fine-tuning. In contrast, some VAE- and diffusion-based FL methods can be designed to reduce communication by sharing compact representations, selected parameters, or generated knowledge instead of full datasets or full models. Recent studies on communication-efficient federated diffusion learning show that diffusion-based strategies can reduce communication cost while maintaining model performance~\cite{vora2024feddm,ahn2022communication}. Future FL-based IDS studies should evaluate not only detection accuracy, but also communication cost, client computation cost, and scalability.

The fourth research direction is realistic federated IDS data generation and benchmarking. As discussed in Section~\ref{subsec:dataset-ids}, realistic FL-based IDS datasets remain limited. Most FL-based IDS studies use centrally collected datasets and partition them artificially across clients. Although this strategy is convenient, it may not reflect real client-level heterogeneity, network topology, temporal changes, or organization-specific attack patterns. FLNET~\cite{kumar2023flnet2023} provides an important step toward FL-oriented IDS evaluation, but more datasets are needed to represent diverse federated deployment scenarios. Generative AI may help create topology-aware and client-specific traffic data, but the generated data must be evaluated for both statistical similarity and network-level validity.

The fifth research direction is LLM-assisted FL-based IDS. To the best of our knowledge, direct studies on LLM-embedded FL-based IDS are still limited. However, LLMs may support FL-based IDS through domain-informed data generation, alert explanation, attack-behavior summarization, traffic-log analysis, and response recommendation. Compared with VAEs, GANs, and diffusion models, LLMs may be better suited for incorporating textual domain knowledge, such as protocol descriptions, network configurations, and attack procedures. However, most pre-trained LLMs are general-purpose models and are not optimized for network-security data. Future studies should investigate network-domain-specific LLMs, parameter-efficient adaptation, and methods for grounding LLM outputs in valid traffic behavior and verified security evidence.

Overall, generative AI-embedded FL-based IDS should be evaluated from multiple perspectives, including detection performance, robustness to adversarial generation, privacy protection, communication efficiency, and realism of generated traffic. Future research should move beyond using generative models only as data generators and study how they can be safely integrated into distributed IDS training and deployment.

\section{Conclusion} \label{sec:conclusion}
% The conclusion goes here.
This survey examined the intersection of generative AI, Federated Learning (FL), and Intrusion Detection Systems (IDSs). We reviewed representative IDS research directions—adversarial machine learning, anomaly-based detection, IoT-oriented IDS, explainable IDS, and benchmark datasets—organized generative AI applications in IDS by four model families (VAEs, GANs, diffusion models, and LLMs) and task objectives, and analyzed how these models are being integrated into FL-based IDS. We now return to the four research questions posed in Section I-A and summarize what the reviewed literature indicates.

RQ1 (How are generative models used to improve IDS?) The reviewed studies show that generative models address IDS along two complementary axes: model family and task objective. VAEs are used primarily for reconstruction-based anomaly detection, latent representation learning, data generation, and augmentation. GANs are most often applied to tabular flow generation, class-imbalance mitigation, data imputation, and, on the offensive side, adversarial traffic generation. Diffusion models are an emerging alternative for synthetic data generation, adversarial purification, and robustness improvement, and tend to offer more stable training than GANs at higher sampling cost. LLMs introduce capabilities the other families lack, including traffic-log analysis, natural-language alert explanation, and knowledge-guided tabular generation. These families also differ in their assumptions: VAEs and GANs operate on fixed feature representations, diffusion models trade computation for stability, and LLMs require traffic to be encoded into model-readable formats. Across all families, the dominant motivation is the same: compensating for limited, imbalanced, incomplete, or non-shareable IDS data.

RQ2 (How reliable is synthetic network traffic for IDS training and evaluation?) The literature indicates that statistical fidelity is necessary but not sufficient. Distribution-level metrics such as KL and JS divergence, Wasserstein distance, FID, and MMD can quantify similarity to real data, but high similarity does not guarantee that generated traffic preserves protocol constraints, temporal dependencies, attack semantics, or network-topology relationships. Several studies report that synthetic data improves some downstream detectors while degrading others, and that low-quality samples can distort decision boundaries. We therefore conclude that synthetic IDS traffic remains only conditionally reliable, and that IDS-specific evaluation—measuring both statistical realism and network-level validity—is still an open requirement rather than a solved problem.

RQ3 (How can generative AI support privacy-preserving and communication-efficient FL-based IDS?) The reviewed FL studies show that generative models can augment local datasets, generate minority-class samples, reduce non-IID skew, and produce privacy-preserving synthetic data without sharing raw traffic. Autoencoder- and VAE-based methods additionally support communication reduction through compact latent representations, and diffusion-based federated methods have shown that high-quality generation can be compatible with reduced communication cost. At the same time, this capability is dual-use: GAN- and diffusion-generated traffic can poison local updates or evade detection, and because the server observes only model updates, such manipulation is harder to detect in FL than in centralized settings. The integration of generative AI into FL-based IDS is thus promising but still early, supported by relatively few primary studies and an open need for robust aggregation and communication-efficient generative sharing.

RQ4 (What datasets and benchmarks are available?) We surveyed widely used IDS datasets (Table II), spanning traditional network intrusion detection, DDoS, IoT, and in-vehicle settings. Most are centrally collected and were not designed to capture client-level heterogeneity, network topology, temporal drift, or organization-specific attack patterns. FLNET2023 is, to our knowledge, the principal dataset that explicitly supports federated evaluation. We therefore conclude that current benchmarks are adequate for centralized generative IDS research but insufficient for federated and topology-aware evaluation, which is one of the most concrete gaps the field faces.

Taken together, these answers point to a small set of priorities for future work. First, more realistic and explicitly federated IDS benchmarks are needed, since artificially partitioned centralized datasets do not reflect real deployment. Second, synthetic traffic generation should become protocol- and topology-aware, and should be evaluated for network-level validity rather than statistical similarity alone. Third, communication-efficient generative FL methods are required so that the cost of sharing generative models does not offset the privacy and efficiency benefits of FL. Fourth, network-domain-specific LLMs, adapted through parameter-efficient tuning and grounded in verified traffic behavior, deserve focused study for IDS data generation, alert explanation, and response recommendation. Finally, because generative models are inherently dual-use, FL-based IDS should be evaluated not only on detection accuracy but also on robustness to adversarial generation, privacy leakage, and communication cost. Building reliable generative AI-enabled IDS will ultimately require combining accurate detection, realistic data generation, privacy preservation, communication efficiency, and security-aware evaluation within a single framework.

\section*{Acknowledgment}
This work was supported by the DEVCOM Analysis Center under Cooperative Agreement Nos. W911QX23D0009 and W911NF2220001, and in part by NSF awards 2148358, 2417062, and 1914635.

\ifCLASSOPTIONcaptionsoff
  \newpage
\fi

\bibliographystyle{unsrt}
\bibliography{IDS_ref,Diffusion_ref,FL_ref,GAN_ref,LLM_ref,VAE_ref, challenges_and_future_direction}

\end{document}